\documentclass[twocolumn,superscriptaddress,10pt]{revtex4-1}
\usepackage{amsmath, amsfonts, amssymb, graphicx, xcolor, float,lineno, multirow,makecell,pifont}
\usepackage[T1]{fontenc} 

\graphicspath{{./figures/}{./si_figures/}}

\renewcommand*{\arraystretch}{1.75}

\newcommand{\TM}{\color{black}}
\newcommand{\AWO}{\color{black}}

\newcommand{\equal}{Corresponding authors}

\newcommand{\GA}{\color{black}}

\usepackage{ifthen}
\newboolean{elife}
\setboolean{elife}{false}

\newcommand{\tick}{{\color{green}\checkmark}}
\newcommand{\cross}{\color{red}\ding{55}}

\begin{document}
\affiliation{
  Laboratoire de Physique de l'\'Ecole normale sup\'erieure, CNRS, PSL University,\\
  Sorbonne Universit\'e, and Universit\'e Paris Cit\'e, Paris, France}
\author{Gabriel~Ath\`enes}
\affiliation{
  Laboratoire de Physique de l'\'Ecole normale sup\'erieure, CNRS, PSL University,\\
  Sorbonne Universit\'e, and Universit\'e Paris Cit\'e, Paris, France}
\affiliation{Bio-Rad SAS, 3 Boulevard Raymond Poincaré, 92430, France}
\author{Adam Woolfe}
\thanks{\equal}
\affiliation{Bio-Rad SAS, 3 Boulevard Raymond Poincaré, 92430, France}
\author{Thierry~Mora}
\thanks{\equal}
\affiliation{
  Laboratoire de Physique de l'\'Ecole normale sup\'erieure, CNRS, PSL University,\\
  Sorbonne Universit\'e, and Universit\'e Paris Cit\'e, Paris, France}
\author{Aleksandra~M.~Walczak}
\thanks{\equal}
\affiliation{
  Laboratoire de Physique de l'\'Ecole normale sup\'erieure, CNRS, PSL University,\\
  Sorbonne Universit\'e, and Universit\'e Paris Cit\'e, Paris, France}

\newcommand{\deftitle}{{Paraplume: A fast and accurate paratope prediction method provides insights into repertoire-scale binding dynamics}}

\title{\deftitle}

\begin{abstract}
The specific region of an antibody responsible for binding to an antigen, known as the paratope, is essential for immune recognition.  Accurate identification of this small yet critical region can  accelerate the development of therapeutic antibodies.  Determining paratope locations typically relies on modeling the antibody structure, which is computationally intensive and difficult to scale across large antibody repertoires. We introduce Paraplume, a sequence-based paratope prediction method that leverages embeddings from protein language models (PLMs), without the need for structural input and achieves superior performance across multiple benchmarks compared to current methods. In addition, reweighting PLM embeddings using Paraplume predictions yields more informative sequence representations, improving downstream tasks such as affinity prediction, binder classification, and epitope binning. Applied to large antibody repertoires, Paraplume reveals that antigen-specific somatic hypermutations are associated with larger paratopes, suggesting a potential mechanism for affinity enhancement. Our findings position PLM-based paratope prediction as a powerful, scalable alternative to structure-dependent approaches, opening new avenues for understanding antibody evolution.

\end{abstract}

\maketitle
\section{Introduction}

\begin{figure*}[t]
\includegraphics[width=1\textwidth]{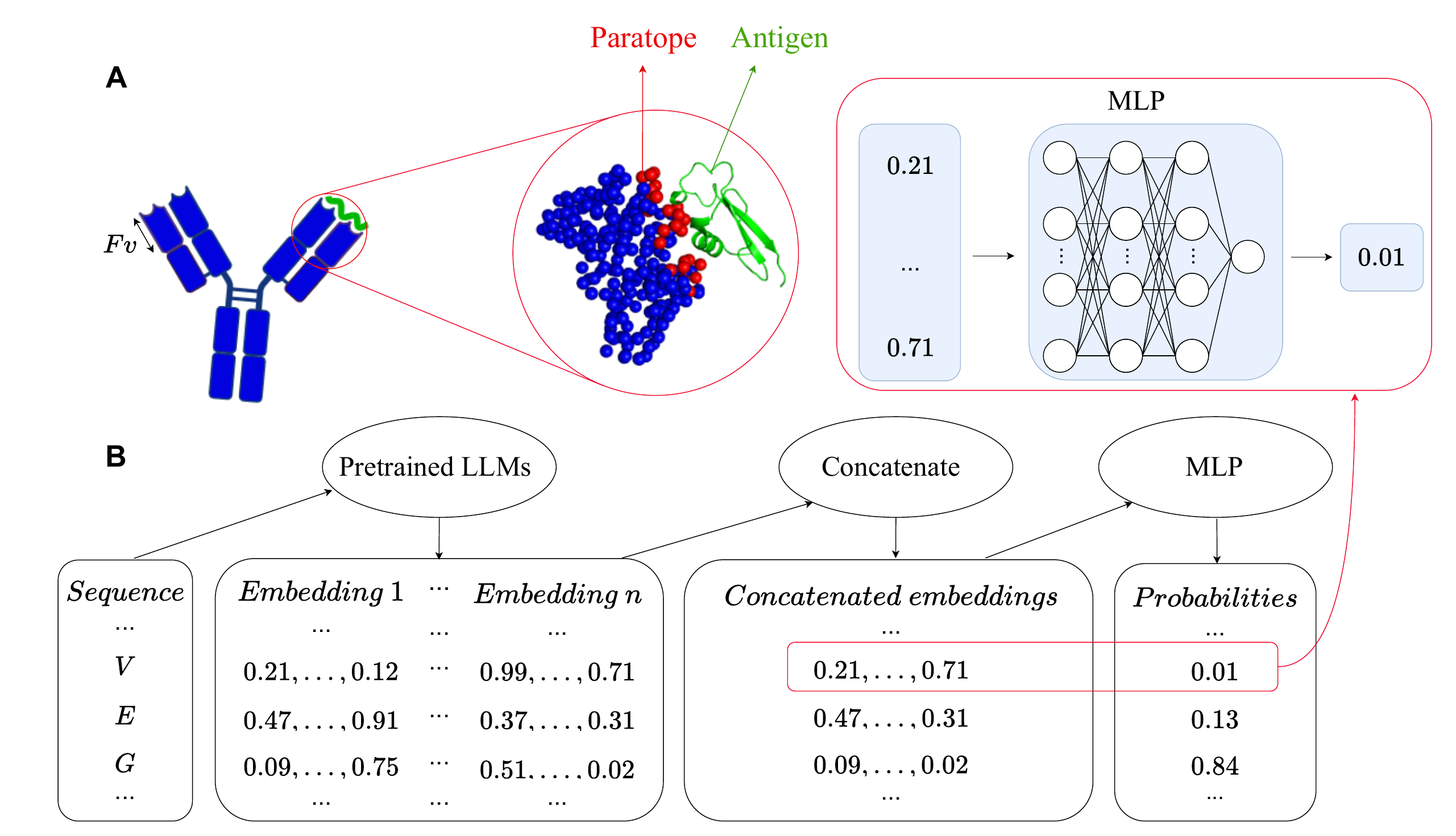} 
\caption{(A) Antibody (blue) binding to an antigen (green), illustrated using the structure of the variable domain ($F_v$ region) of the mouse anti-lysozyme antibody 1BVK. Amino acids are represented using carbon alpha $C_{\alpha}$ atoms, and the paratope is colored red. Amino acids are labeled as belonging to the  paratope if any non-hydrogen atom is within a distance of 4.5 \r{A} of a non-hydrogen antigen atom. (B) The pipeline used for paratope prediction. The antibody sequence is given as input to protein language models (PLMs), the last embedding layer of which is  concatenated and fed to a multi objective multilayer perceptron (MLP). The MLP calculates probabilities of amino acids belonging to a paratope. 
}
\label{fig1}  
\end{figure*}

{\GA Antibodies are specialized proteins of the immune system, produced by B cells, that recognize foreign pathogens, either neutralizing them directly or marking them for removal. This highly specific recognition is determined by the antibody's variable regions and is refined through a Darwinian process known as affinity maturation, which B cells undergo after encountering an antigen. During this process, the genes encoding the variable regions undergo somatic hypermutation, and B cells producing higher-affinity antibodies are selectively expanded.
The paratope comprises specific amino acids in the variable regions of the antibody that directly interact with residues on the target antigen, known as epitopes, upon binding (Fig.~\ref{fig1}A). This interaction determines the antibody's binding specificity and affinity, both of which are essential for an effective immune response.}
Mapping the specific location of the paratope has important applications in biotechnology and medicine, especially in the design of therapeutic antibodies, as accurate predictions of antibody binding sites can help identify key residues for targeted mutations that modify binding properties \cite{rabia2018understanding} {or that should be avoided during engineering of antibodies for enhanced developability.}

However, experimental methods for determining antibody-antigen binding interactions are slow and resource-intensive \cite{wouters2020estimated}. In contrast, computational methods such as molecular docking have been developed as a more efficient alternative, offering faster, lower-cost approaches to predict how antibodies and antigens can bind~\cite{fan2019progress}. While promising, these tools still face limitations in accuracy \cite{ambrosetti2020modeling}, especially at a scale required for high-throughput applications \cite{bender2021practical}, and they require the 3D structures of both antibodies and antigens.
Although the recent release of Alphafold 3~\cite{Abramson2024} shows an improvement in modelling accuracy of the antibody-antigen complex, it is limited in the antibody-docking task \cite{hitawala2024has} and requires the antigen sequence. 

To address these challenges, numerous methods have been developed for predicting antibody paratopes. Parapred \cite{liberis2018parapred}, a freely accessible sequence-based tool, uses convolutional neural networks to extract local sequence features and recurrent neural networks to capture long-range dependencies. 
While practical, Parapred is limited to predicting paratopes within the complementarity-determining regions (CDRs) of the antibody, requiring sequence numbering as a prerequisite. These 6 CDRs (three in the heavy chain and three in the light chain) encompass the majority of the paratope, thereby simplifying the training of supervised models.
However, recent advancements in methods that leverage antibody 3D structural information have surpassed Parapred in performance, leading state-of-the-art paratope prediction approaches to predominantly rely on either experimentally determined structures or high-quality modeled counterparts. Paragraph~\cite{chinery2023paragraph} models the 3D antibody structure using AbodyBuilder~\cite{leem2016abodybuilder} and Ablooper~\cite{abanades2022ablooper}, represents the structure as a graph based on amino acid distances, and then uses equivariant graph neural networks~\cite{satorras2021n} for the paratope prediction task.
Similarly, methods like PECAN~\cite{pittala2020learning} and MIPE~\cite{wang2024improving} {\GA require the 3D structures of both the antibody and antigen to predict the paratope.}
However, {\GA experimentally determined} 3D structures are not always readily available, and generating accurate 3D models introduces additional challenges. This reliance on structure prediction models not only leads to a significant drop in performance, as observed in~\cite{wang2024improving}, but also requires time-intensive pre-processing steps to compute interacting residues. 
{\GA These limitations underscore the need for more precise and scalable computational models that can effectively identify antibody binding sites.}

{\GA Powered by the Transformer architecture, protein language models (PLMs) pretrained on huge databases of protein sequences have been applied to tasks such as secondary structure prediction and contact map estimation \cite{verkuil2022language, elnaggar2021prottrans}. 
Their ability to extract structural and functional information from sequence data alone makes them especially valuable for predicting antibody binding sites in the absence of structural information.}
We introduce Paraplume, a sequence-based, antigen-agnostic paratope inference method that overcomes data limitations by leveraging embeddings from six PLMs and achieves state-of-the-art performance on three independent datasets. 
{The speed and accuracy of Paraplume now {\GA enable applications to large antibody repertoire sequencing datasets, which were limited by the} computational constraints of prior methods. 
{\GA Using Paraplume,} we compared naive and antigen-exposed antibody repertoires {\GA and identified a clear signal of paratope evolution.}}

\section{Results}

\subsection{Paraplume}
Paratope prediction consists of assigning a label 1 to an amino acid if it belongs to the paratope and 0 otherwise. Supervised methods construct training and testing datasets by annotating amino acids with paratope labels using the {\GA experimentally determined} 3D structures of antibody-antigen complexes available in SabDab~\cite{dunbar2014sabdab}. Concretely, an antibody amino acid is labeled 1 if at least one of its non-hydrogen atoms is within $4.5 \ \text{\r{A}}$ of a non-hydrogen atom of the antigen. 

{\GA The main challenge in using structural data to train supervised models for paratope inference is the limited availability of structures in SAbDab.
To mitigate this issue, we leverage information from millions of sequences by representing all amino acids from the variable region as embeddings derived from protein language models (PLMs).
PLMs are typically trained in an unsupervised manner on large protein sequence datasets. Antibody-specific PLMs are either trained directly on large collections of antibody sequences or adapted from general protein PLMs through finetuning.
These models produce embedding vectors that} {contain information not just about the amino acid itself, but also about its sequence context through the attention mechanism.}
{\GA While most approaches using PLMs rely on a single model, we hypothesized that concatenating embeddings from multiple PLMs could provide complementary information not captured by any individual model alone.}
Specifically, each amino acid is represented as the concatenation of embeddings from six language models: AbLang2\cite{olsen2024addressing}, Antiberty\cite{ruffolo2021deciphering}, ESM-2\cite{verkuil2022language}, IgT5\cite{kenlay2024large}, IgBert\cite{kenlay2024large}, and ProtTrans~\cite{elnaggar2021prottrans}. 
These concatenated embeddings are then input to a Multi-Layer Perceptron {\GA (MLP)} that uses paratope labels for training (Figure \ref{fig1}B).
{\GA A detailed discussion of the model's design choices is available in Section~\ref{IV.A}.}
Once the embeddings are computed, training the model becomes computationally feasible using only a CPU. {Here, we introduce Paraplume, the resulting sequence-based supervised paratope prediction model (Figure 1B).} Paraplume assigns a probability to each amino acid in the input sequence, reflecting its likelihood of being part of the paratope. It is trained by minimizing the Binary Cross-Entropy loss between the predicted probabilities and the true labels (cf. Section~\ref{IV.B}). Although the three-dimensional structure is essential for generating the labels used during training, Paraplume does not require structural information to make predictions. Paraplume takes as input either the heavy chain, the light chain, or paired heavy and light chains, and makes predictions solely based on sequence data {\GA (cf. Section~\ref{IV.C})}. 
Paraplume is also antigen-agnostic, meaning it does not require any antigen-specific information for its predictions.
A key advantage of Paraplume's sequence-based design is its computational efficiency, allowing paratope predictions for 1000 sequences in 3 minutes (50 seconds if only using one ESM embedding) using a single GPU (cf. Fig.~\ref{si:fig1}),
facilitating large-scale analysis of antibody sequence {repertoires}.

\subsection{{\GA Performance comparison}}
\begin{table*}[t]
	\centering
	{\large Using sequences as inputs}
	\huge
	\renewcommand{\arraystretch}{1.4} 
	\setlength{\tabcolsep}{12pt} 
	\resizebox{\textwidth}{!}{%
		\begin{tabular}{l|cccc|cccc|cccc|cc}
			\hline
			{\multirow{2}{*}{\bf Model}}          & \multicolumn{4}{c|}{\textbf{PECAN Dataset}} & \multicolumn{4}{c|}{\textbf{Paragraph Dataset}} & \multicolumn{4}{c|}{\textbf{Mipe Dataset}} & {\multirow{2}{*}{\makecell[c]{\bf Structure \\ \bf modeling free}}} & {\multirow{2}{*}{\makecell[c]{\bf Antigen \\ \bf agnostic}}} \\
			\cline{2-13}
			& {\bf PR} & {\bf ROC}   & {\bf F1} & {\bf MCC} &{\bf PR} & {\bf ROC}   & {\bf F1} & {\bf MCC} &{\bf PR} & {\bf ROC} & {\bf F1} & {\bf MCC} &  \\
			\hline\hline
			Baseline           & 0.626             & \underline{0.952}           &  0.665     &    0.635    & 0.624          & \underline{0.952}          &  0.622    &    0.654    & 0.465                  &  0.931            &  0.536            & 0.177          & \tick & \tick \\
			Parapred           & 0.646             & 0.930           &  -         &    -        & -              & -              &  -        &    -        & 0.652                  & 0.868             &  -                & 0.503          & \tick & \tick \\
			Paragraph  (ABB)   & \underline{0.696} & 0.934           &  \bf 0.685 &   \underline{0.654}     & \underline{0.725}          & 0.934          &  \underline{0.696}    &    \underline{0.669}    & 0.689                  & \underline{0.937}             &  \underline{0.617}            & \underline{0.596}          & \cross & \tick  \\
			PECAN  (ABB)       & 0.675             & \underline{0.952}           &  -         &    -        & -              & -              &  -        &    -        & -                      & -                 & -                 &  -             & \cross & \cross \\
			MIPE \ (AF2)       & -                 & -               &  -         &    -        & -              & -              &  -        &    -        &  \bf 0.723             & 0.910             &  \underline{0.617}            & 0.531          & \cross & \cross  \\
			\hline
			Paraplume          & \textbf{0.730}    & \textbf{0.963}  & \underline{0.682}      &   \bf 0.657 & \textbf{0.758} & \textbf{0.966} & \bf 0.701 & \bf 0.676   & \underline{0.716}      & \textbf{0.962}    & \bf 0.651         & \textbf{0.632}         & \tick & \tick \\
		\end{tabular}
	}
	\caption{ Comparison of methods that use sequences as inputs. Paragraph and PECAN model the 3D structures from the sequences with ABodyBuilder (ABB) \cite{leem2016abodybuilder}, while MIPE uses AlphaFold2 (AF2), and requires both antibody and antigen sequences. All other methods operate directly on sequences without requiring structural modeling. Performance metrics (PR AUC, ROC AUC, F1 score, and MCC) with additional model characteristics (structure modeling free and antigen agnostic) for models evaluated on PECAN, PARAGRAPH, and MIPE datasets. The highest value in each column is in bold, the second best is underlined.}
	\label{tab1}
      \end{table*}

\begin{figure*}[t]
	\includegraphics[width=1\textwidth]{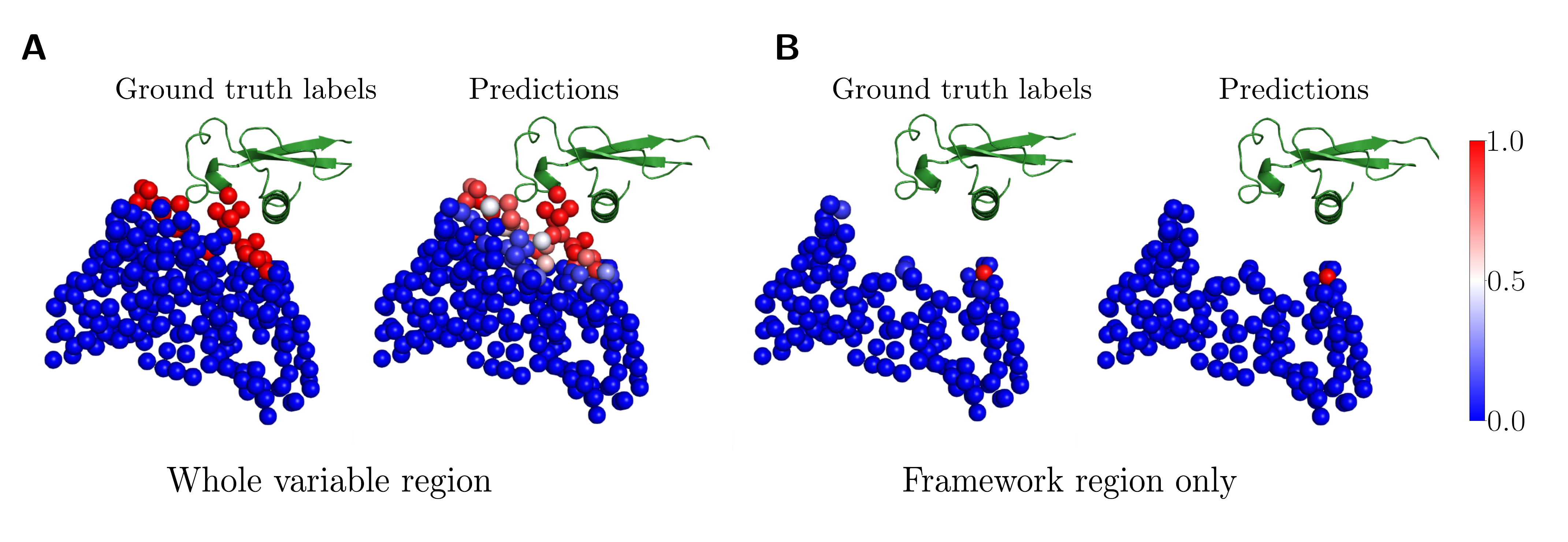} 
	\caption{
		 (A) Comparison of ground truth paratope labels (left) and Paraplume model predictions (right) for the full variable region of the 6B0S antibody-antigen complex, which was not included in the training set. For visualization, antibodies were depicted as spheres and the antigen in a cartoon representation (green) in PyMOL~\cite{delano2002pymol}.  In the ground truth structure, residues forming the paratope are highlighted in red. The colorbar shows the probability of a given amino acid being a paratope residue.  For clarity, only the $C_{\alpha}$ carbon of each residue is depicted.  (B) Same structure as in (A) but restricted to amino acids belonging to the framework region.} 
\label{fig2}  
\end{figure*}

We evaluate the performance of Paraplume 
in comparison to existing paratope prediction methods across three datasets using four evaluation metrics. 
The PECAN dataset comprises 460 antibody-antigen complexes with paired heavy and light chains, all resolved at sub-3~\r{A} resolution, and is divided into 205 training, 103 validation, and 152 test samples. The Paragraph dataset, extracted from the Structural Antibody Database (SAbDab) as of March 31, 2022, consists of 1,086 antibody-antigen complexes, partitioned into training, validation, and test sets in a 60-20-20\% split. The MIPE dataset includes 626 antibody-antigen complexes, with 90\% allocated for training and 10\% for testing.

Paraplume is underlied by several choices of architecture and hyperparameters, which are justified and discussed in detail in Section~\ref{IV.A}. The results presented below were obtained for the best performing model. All benchmark evaluations are done with identical hyperparameters and modeling choices, without tuning on individual datasets.
Model performance is assessed using four metrics: the precision-recall area under the curve (PR AUC) and the receiver operating characteristic area under the curve (ROC AUC), which evaluate classification performance in imbalanced datasets; the F1-score (F1), representing the harmonic mean of precision and recall; and the Matthews correlation coefficient (MCC). Both F1 and MCC are computed using the standard 0.5 threshold to binarize predictions. Following the approach used for other methods, each metric was averaged over all proteins in the test set.

The benchmarked methods vary significantly in their approaches: some predict paratopes directly from sequence data (Parapred, Paraplume), others rely on modeling the 3D structure from the sequence (Paragraph, PECAN, MIPE)---a preprocessing step that reduces scalability---and some make predictions based on the experimentally determined structure of the antibody alone or in combination with the antigen (Parasurf-Fv, and versions of Paragraph, PECAN and MIPE). Because the experimentally determined antibody structures used for training and testing by these methods are derived from antibody-antigen complexes, each structure serves both as model input and for paratope labeling. Given that antigen binding can induce substantial conformational changes in antibodies~\cite{sela2012systematic}, this raises concerns about the generalizability of such models to unbound antibody structures.
To ensure a fair comparison, in Table \ref{tab1} we compare Paraplume with methods that do not take experimentally determined structures as input: Parapred, PECAN, Paragraph, MIPE, and the baseline method described in \cite{chinery2023paragraph}. Paragraph and PECAN use ABodyBuilder \cite{leem2016abodybuilder} for structure modeling from the sequence, while MIPE uses AlphaFold2 \cite{jumper2021highly}. 
In contrast, Parapred and Paraplume do not require structure modeling. Additionally, MIPE and PECAN incorporate antigen information in their predictions.
\begin{table*}[t]
	\centering
	{\large Using experimentally determined structures as inputs}
	\huge
	\renewcommand{\arraystretch}{1.4} 
	\setlength{\tabcolsep}{12pt} 
	\resizebox{\textwidth}{!}{%
		\begin{tabular}{l|cccc|cccc|cccc|c}
			\hline
			{\multirow{2}{*}{\bf Model}}          & \multicolumn{4}{c|}{\textbf{PECAN Dataset}} & \multicolumn{4}{c|}{\textbf{Paragraph Dataset}} & \multicolumn{4}{c|}{\textbf{Mipe Dataset}} &  {\multirow{2}{*}{\makecell[c]{\bf Antigen \\ \bf agnostic}}}  \\
			\cline{2-13}
			& {\bf PR} & {\bf ROC} & {\bf F1} & {\bf MCC} & {\bf PR} & {\bf ROC} & {\bf F1} & {\bf MCC} & {\bf PR} & {\bf ROC} & {\bf F1} & {\bf MCC} \\
			\hline\hline
			Paragraph        & \underline{0.754}        & 0.940    & \bf{0.703}  &  0.674   & 0.770        & 0.939    & \bf{0.719}  & \bf{0.692}   & 0.742       & 0.943    & 0.651 & 0.634  & \tick \\
			Parasurf-Fv      & 0.733        & \underline{0.955}    & 0.647  & 0.612   & \textbf{0.793} & \underline{0.967}    & 0.698  & 0.676   & \bf 0.781       & \textbf{0.967}    & \bf 0.690  & \textbf{0.659}  & \tick \\
			PECAN      & 0.700        & \underline{0.955}    & -  & -   & -            & -        & -  & -   & 0.713       & 0.915    & -  & 0.558  & \cross \\
			MIPE      & -            & -        & -  & -   & -            & -        & -  & -   & 0.741 & 0.927    & 0.627  & 0.554  & \cross \\
			\hline
			Paraplume-G          & \textbf{0.772} & \textbf{0.965}  &  \underline{0.697}  & \bf 0.675   & \underline{0.791} & \textbf{0.968}  &  \underline{0.704}  & \underline{0.683}   & \underline{0.753}       &  \underline{0.964}    & \underline{0.663} & \underline{0.648}  & \tick \\
		\end{tabular}
	}
	\caption{Comparison of methods that use experimentally determined structures as inputs. Performance metrics (PR AUC, ROC AUC, F1 score, and MCC) with additional model characteristics (antigen agnostic) for models evaluated on PECAN, PARAGRAPH, and MIPE datasets. The highest value in each column is in bold, the second best is underlined.}
	\label{tab2}
\end{table*}
For the PECAN and Paragraph datasets, performance metrics for all methods were obtained from \cite{chinery2023paragraph}. For the MIPE dataset, results were taken from \cite{wang2024improving}, with the exception of Paragraph. 
To present Paragraph in the most favorable light, we retrained the model and evaluated its performance using inputs generated from structures predicted by ABodyBuilder3 (ABB3) \cite{kenlay2024abodybuilder3}, a state-of-the-art structure prediction model. Note that this method may slightly overestimate Paragraph's performance, as some sequences in the MIPE test set are also in the ABB3 training data.
As with other methods, Paraplume was trained and evaluated separately on each of the three datasets using their respective predefined splits.
For the PECAN and Paragraph datasets, Paraplume was trained using 16 random seeds. For each seed and dataset, early stopping was applied by retaining the model weights that achieved the highest PR AUC on the validation set, and performance was then evaluated on the corresponding test set.
On the MIPE dataset, we performed a 5-fold cross-validation on the training-validation set, consistent with other methods \cite{wang2024improving} and \cite{papadopoulos2025parasurf}. 
For each fold we trained our model on the training set, retained the weights that maximized the PR-AUC on the validation set for testing on the independent test set.
The reported results are averaged over the 5 folds and 5 seeds as done in \cite{wang2024improving}.
Using only antibody sequence information, Paraplume outperformed all other methods across all four evaluation metrics on the Paragraph datasets, and for three out of four metrics for the PECAN and MIPE datasets {(Table \ref{tab1})}.

We show an example of Paraplume's predictions, trained on Paragraph's train set, compared to the ground truth labels of {an antibody specific to} the aTSR domain of a circumsporozoite protein (PDB: 6B0S) from Paragraph's test set (Figure~\ref{fig2}A).
{\GA Paraplume correctly identified all 23 experimentally determined paratope residues (TPR = 100\%) while falsely labeling 9 of 205 non-paratope residues (FPR $\sim$ 4.4\%).}
Paraplume successfully predicts  paratope residues located in the framework region (Figure~\ref{fig2}B), which is not achievable with methods limited to predictions within the CDR $\pm 2$ region such as Paragraph {or Parapred}.

{\GA To better understand the contribution of each of the six PLMs used in Paraplume, and to explore whether a more lightweight variant could retain strong performance, we conducted an ablation study. We evaluated model performance using all embeddings, individual embeddings, or all but one (Table~\ref{si:tab1}). 
While no single configuration consistently outperformed others across all datasets, using all six embeddings generally yielded more robust performance.
However, using only the ESM embedding achieved strong results with significantly reduced computational cost, motivating the development of Paraplume-S, a smaller and faster variant of Paraplume.
A comparison of inference-time computational costs and $\text{CO}_2$ equivalent emissions for Paraplume, Paraplume-S, and Paragraph under both GPU and CPU settings is shown in Figure~\ref{si:fig1}.

Finally, since paired chain data is often unavailable in large-scale bulk sequencing studies, it is important to assess whether Paraplume maintains reliable performance on single-chain sequences. To this end, we evaluated Paraplume on single-chain variants (cf. Section~\ref{IV.C} for details) and observed only a minor decrease in performance, supporting its applicability to heavy chain only repertoires (Table~\ref{si:tab2}).}

\subsection{Combining Paraplume and Paragraph for experimentally determined structures}
Recent studies \cite{chinery2023paragraph, wang2024improving} have demonstrated notable differences in performance between models using experimentally determined structures and those relying on predicted structural models. Among methods that use experimentally determined structures (Table~\ref{tab2}), some such as Paragraph, show improved performance within the CDR regions compared to Paraplume, but this advantage is lost in the framework regions, or when using modeled structures instead of experimentally determined structures (Table~\ref{si:tab3}). 
This could be because Paragraph is trained in the CDR$\pm 2$ region, where the paratope-to-non-paratope ratio is higher ($1:3$ compared to $1:10$ in the whole variable region), thereby stabilizing training.

To further increase performance across the entire sequence we developed Paraplume-G {(Graph-based Paraplume)}, which uses structural information and combines the strengths of both Paragraph and Paraplume. Specifically, Paragraph, trained using the parameters described in the original paper, is used to predict residues in the CDR$\pm 2$ region, while Paraplume is applied to predict residues outside this region.

Table~\ref{tab2} presents results for methods that rely on experimentally determined structures, comparing Paraplume-G with Paragraph, Parasurf \cite{papadopoulos2025parasurf}, PECAN, MIPE, and the  baseline method described in \cite{chinery2023paragraph}. Performance metrics for Parasurf and MIPE across the three datasets were obtained from their respective publications. For PECAN, results were taken from \cite{chinery2023paragraph} for the PECAN dataset and from \cite{wang2024improving} for the MIPE dataset.
For Paragraph, we used the results on the PECAN and Paragraph datasets from \cite{chinery2023paragraph} and retrained Paragraph using experimentally determined structures for the MIPE dataset, following the approach described in \cite{chinery2023paragraph}, averaging the results across 16 different seeds. We observed significantly higher performance on the MIPE dataset compared to the results reported in \cite{wang2024improving}. Paraplume-G demonstrated performance comparable to state-of-the-art methods for experimentally determined structure-based paratope prediction, ranking first or second across all 12 metrics derived from the three datasets. It outperformed Parasurf on 7 of the 12 comparison points and surpassed Paragraph on 9 of them.

\subsection{Calculating Performance Upper Bounds of Paratope Prediction Using Identical Antibody Arms in Antibody-Antigen structures}
\begin{figure*}[t]
	\includegraphics[width=0.9\textwidth]{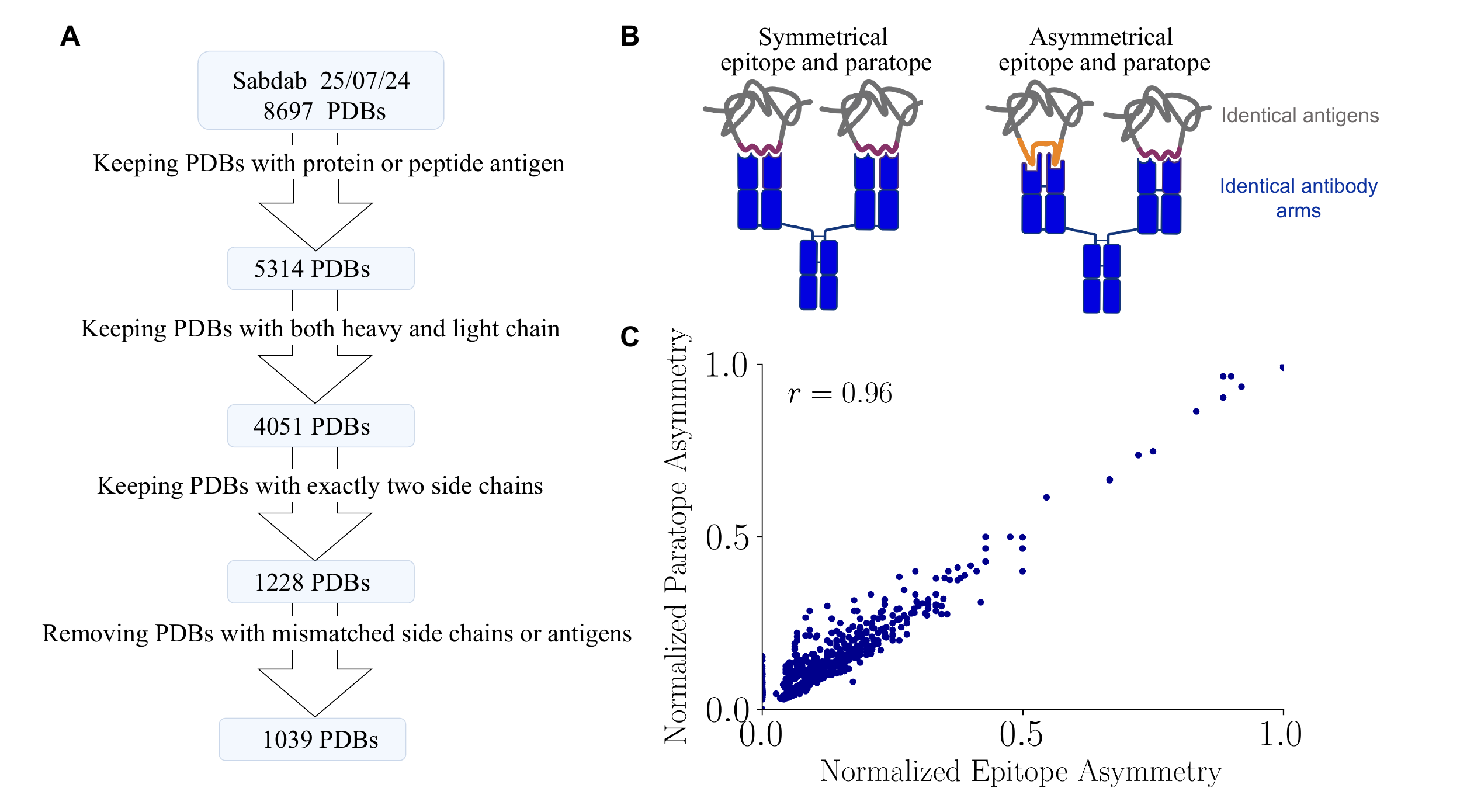}
	\caption{ (A) Dataset curation for the epitope asymmetry analysis, with the number of PDB structures at each stage. 
	(B) Cartoon of antibody-antigen complexes with symmetric and asymmetric paratopes and epitopes. An antibody side chain paratope binds to an epitope on the antigen side chain. In the asymmetric case two identical antibody sequences bind different epitopes, using different paratopes. 
	(C) Normalized paratope asymmetry correlates strongly  with the normalized epitope asymmetry (Pearson correlation coefficient), where each point represents a distinct antibody-antigen complex. 
	}
	\label{fig3}  
\end{figure*}

Proteins are not rigid structures but instead exist as ensembles of conformations that fluctuate over time. A recent study~\cite{fernandez2020antibody} suggests that a single antibody can adopt multiple conformations, underscoring the structural flexibility of CDR loops in the context of antigen recognition.
{This suggests that paratope definition may not be a straightforward problem, setting a potential upper bound of any paratope prediction method. To explore the extent to which paratope definition may vary}, we curated a dataset of 1,039 antibody-antigen complexes from the SabDab database in which a single antibody binds two identical antigens, one for each arm, allowing direct comparison of ground-truth paratope annotations across the two arms (Figure~\ref{fig3}A, see Section~\ref{IV.D} for details). We quantified the variability between the two antibody chains using a metric we define as paratope asymmetry (and analogously epitope asymmetry), which counts the number of residues present in the paratope or epitope in one arm, but not the other
{\GA (see Section~\ref{IV.E} and Figure~\ref{si:fig2}A and E)}.
{\TM We found that paratope and epitopes can vary by more than 10 amino acids between arms}
(Figure~\ref{si:fig2}B and F). To account for size-dependent effects Figure~\ref{si:fig2}C and G), we also define a normalized version of these metrics based on the total paratope or epitope size (see Section~\ref{IV.E} and Figure~\ref{si:fig2}D and H).

We investigated potential sources of asymmetry, such as antigen type, the structure determination method, PDB resolution, and sequence differences from missing residues, but found only weak correlations (Figure~\ref{si:fig3}). By contrast, we observed a strong correlation between normalized paratope and epitope asymmetry (Figure~\ref{fig3}C), indicating that structural changes in the antibody are closely mirrored by changes in the antigen interface. This suggests that the observed asymmetry reflects real biological dynamics rather than technical artifacts. 

This biological variability provides an empirical upper limit on the performance of sequence-based paratope prediction models. For each of the three benchmark test datasets, we extracted the subset of structures present in our curated set comprising 80 Paragraph sequences, 18 MIPE sequences, and 66 Pecan sequences. We measured the F1 score by treating one arm's labels as the ``ground truth'' and the other as ``predictions''. 
Across all three datasets, we found this upper bound to be consistently around 95\% F1 (Table~\ref{tab3}, with our model?s performance included for comparison). 
{\GA To assess how this variability affects our model, we further analyzed its predictions on ambiguous residues, defined as those with discordant paratope labels between the two arms, and on consistent residues, where labels agreed. We observed that predictions for ambiguous residues were more frequently distributed around 0.5, indicating reduced confidence and greater difficulty in predicting paratopes for residues subject to biological variability (Fig.~\ref{si:fig4}).
Together these observations highlight} a critical limitation: even under ideal conditions, where each antibody binds to a single antigen type, perfect prediction remains impossible due to natural structural variability. 
In fact, as antibodies may interact with a diverse range of antigen types, we expect the {maximum} achievable performance to be even lower. To set the corresponding upper bound, one would need to compare 3D structures of the same antibody binding to distinct antigens, and measure the difference in their paratopes. However no such data is available to our knowledge. Thus, how much room there is left for the improvement of computational paratope prediction methods remains an open question.
\begin{table}[h]
	\centering
	\resizebox{\columnwidth}{!}{%
		\begin{tabular}{l|cc|cc|cc}
			\multirow{2}{*}{Method} & \multicolumn{2}{c|}{PECAN} & \multicolumn{2}{c|}{Paragraph} & \multicolumn{2}{c}{Mipe} \\
			& F1 & MCC & F1 & MCC & F1 & MCC \\
			\hline\hline
			Upper Bound & 0.947 & 0.944 & 0.953 & 0.949 & 0.961 & 0.958 \\
			Paraplume   & 0.663 & 0.637 & 0.711 & 0.686 & 0.712 & 0.689 \\
		\end{tabular}
	}
	\caption{F1 score and MCC for paratope prediction for the \textit{Upper Bound} and \textit{Paraplume} conditions across the PECAN, Paragraph, and MIPE datasets.}
	\label{tab3}
\end{table}

\subsection{Application to large scale antibody sequence datasets}

\subsubsection*{Probability of amino acid belonging to a paratope correlates with impact on binding affinity}

To demonstrate the model’s applicability in exploring antibody-antigen binding, we analyzed a dataset from {Phillips et al.}~\cite{phillips2021binding}, comprising antibody sequences with experimentally measured binding affinities to three influenza strains (H1, H3 and FluB).
This dataset was constructed by introducing all possible combinations of 16 mutations that differentiate the broadly neutralizing antibody (bnAb) CR9114 from its germline, totaling $2^{16}$ unique sequences. 
The study revealed that broad neutralization emerges sequentially, with binding initially increasing for the H1 strain, followed by H3, and finally FluB, as mutations accumulate in the germline. 
Moreover, the mutational effects on binding affinity exhibit a nested structure, where antibodies binding to FluB also bind to H3, and those binding to H3 also bind to H1.

\begin{figure*}[t]
	\includegraphics[width=\textwidth]{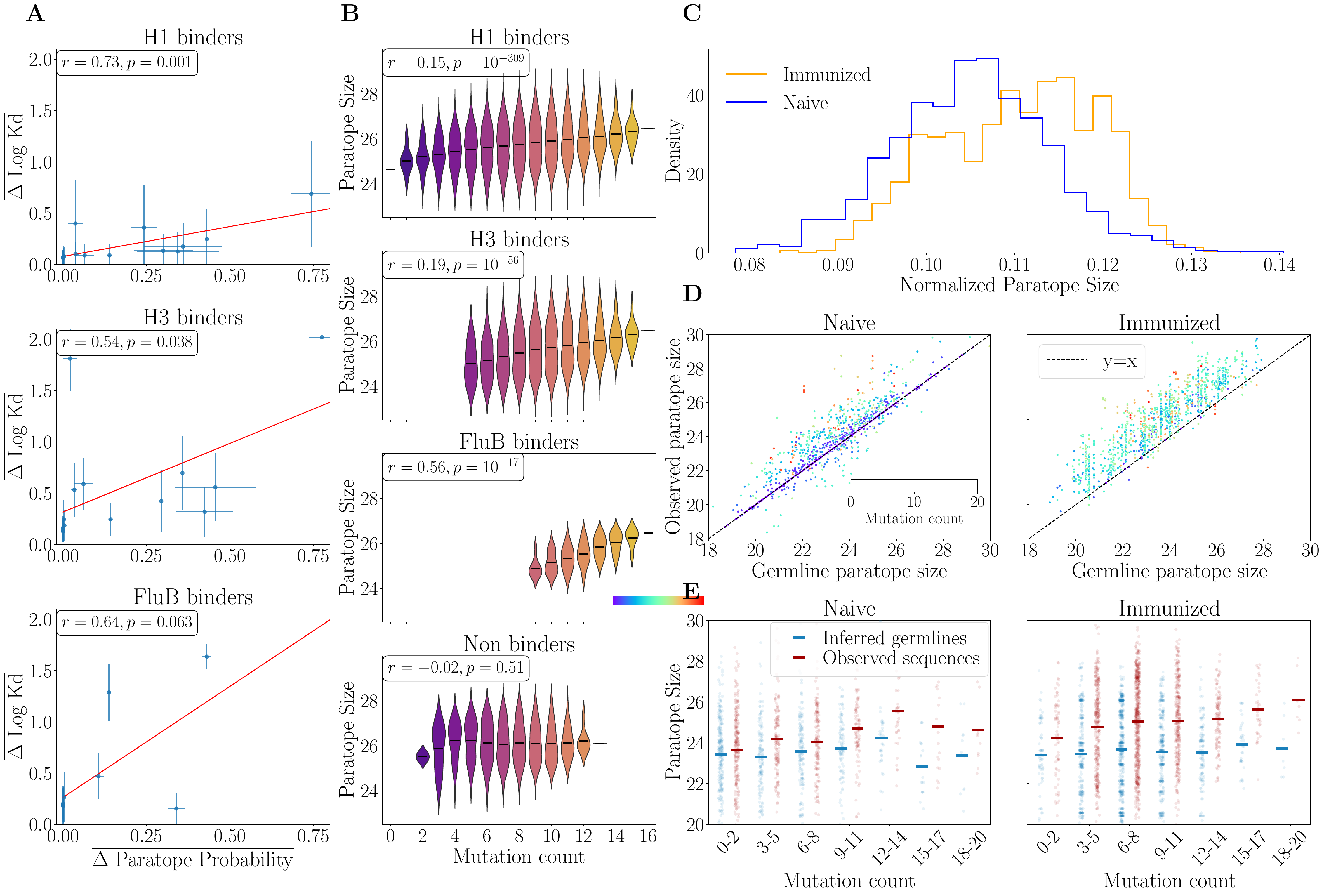}
	\caption{(A) 
		Correlation between the average change in affinity and the average change in the probability for an amino acid to belong to a paratope across the 16 mutated positions of bnAb 9114. {\GA Averages are computed across all antibody variants with measurable affinity in~\cite{phillips2021binding} for each of the H1, H3, and FluB antigens}. (B) {\GA Paratope size as a function of amino acid mutation count for three groups of binders and non-binders, based on experimental affinity measurements from~\cite{phillips2021binding}. Non-binders are defined as sequences with no measurable affinity to any of the three strains.} (C) Normalized paratope size as a function of mutation count for a repertoire of IgG antibodies from mice immunized with tetanus toxoid \cite{gerard2020high} with antibodies sorted for binding to the antigen, compared to naive antibodies from the same mouse species \cite{goldstein2019massively}.
	(D) Comparison of paratope size between antibody sequences and their inferred germline sequences in the antibody repertoires of naive mice (left) and immunized mice (right). (E) Paratope size of observed antibody sequences and their germline sequences across different amino acid mutation count bins for naive mice (left) and immunized mice (right). {\GA The mutation count represents the number of amino acid differences between each antibody sequence and its germline, which is why germline sequences are also assigned mutation counts.}}
	\label{fig4}  
\end{figure*}

To examine the role of the paratope for binding affinity, we used Paraplume, trained on the complete expanded dataset from \cite{chinery2023paragraph}, excluding the 2 PDB structures of the CR9114 bnAb (PDB labels 4FQI and 4FQY), to predict the paratopes of all $2^{16}$ antibody variants.
{\GA For each strain, we excluded antibody sequences that did not exhibit measurable binding affinity to the corresponding antigen ($-\log(K_d) = 7$ for H1 and $-\log(K_d) = 6$ for H3 and FluB), resulting in separate subsets of binders for each strain. Within each subset, and for each of the 16 specific mutations, we identified sequence pairs that differed only by that particular mutation.} For each pair, we computed the absolute difference in the predicted probability of the mutation site belonging to a paratope and the absolute difference in binding affinity {\GA for the strain}.
{\GA For each strain subset we} then averaged these differences across all pairs to obtain the mean absolute difference in probability of the mutation belonging to a paratope, $\overline{\Delta \text{Paratope Probability}}$, and the mean absolute difference in binding affinity, $\overline{\Delta \log K_D}$. As a result we obtain the  average change in the probability that this residue is part of the paratope for each of the 16 mutations, which correlates positively with the average change in the binding affinity, for each strain (Figure~\ref{fig4}A). 
Mutations that result in significant changes in the probability of the amino acid to belong to the paratope suggest that these mutations are likely to influence the binding of the amino acid at that position, thereby affecting affinity.
 
\subsubsection*{Mutations increase paratope size}

We next investigated the impact of mutations on paratope size, computed as the sum of the probabilities of belonging to a paratope for all amino acids in both the heavy and light chains. 
Analysis across all antigens reveals a positive correlation between paratope size and mutation count (Fig.~\ref{fig4}B).
This correlation is absent in non-binding antibodies (Fig.~\ref{fig4}B, bottom panel), implying that those unable to bind strongly to any of the three strains likely failed to develop a corresponding paratope.
However, the interpretation of a computationally identified paratope for {a} non-binding antibody is unclear. Since the model was trained on antibodies with a defined antibody-antigen complex, it might be biased, resulting in overestimated paratope predictions for antibodies that do not interact with an antigen.

\subsubsection*{Validation on whole antibody repertoire}
\begin{figure*}[t]
	\includegraphics[width=0.9\textwidth]{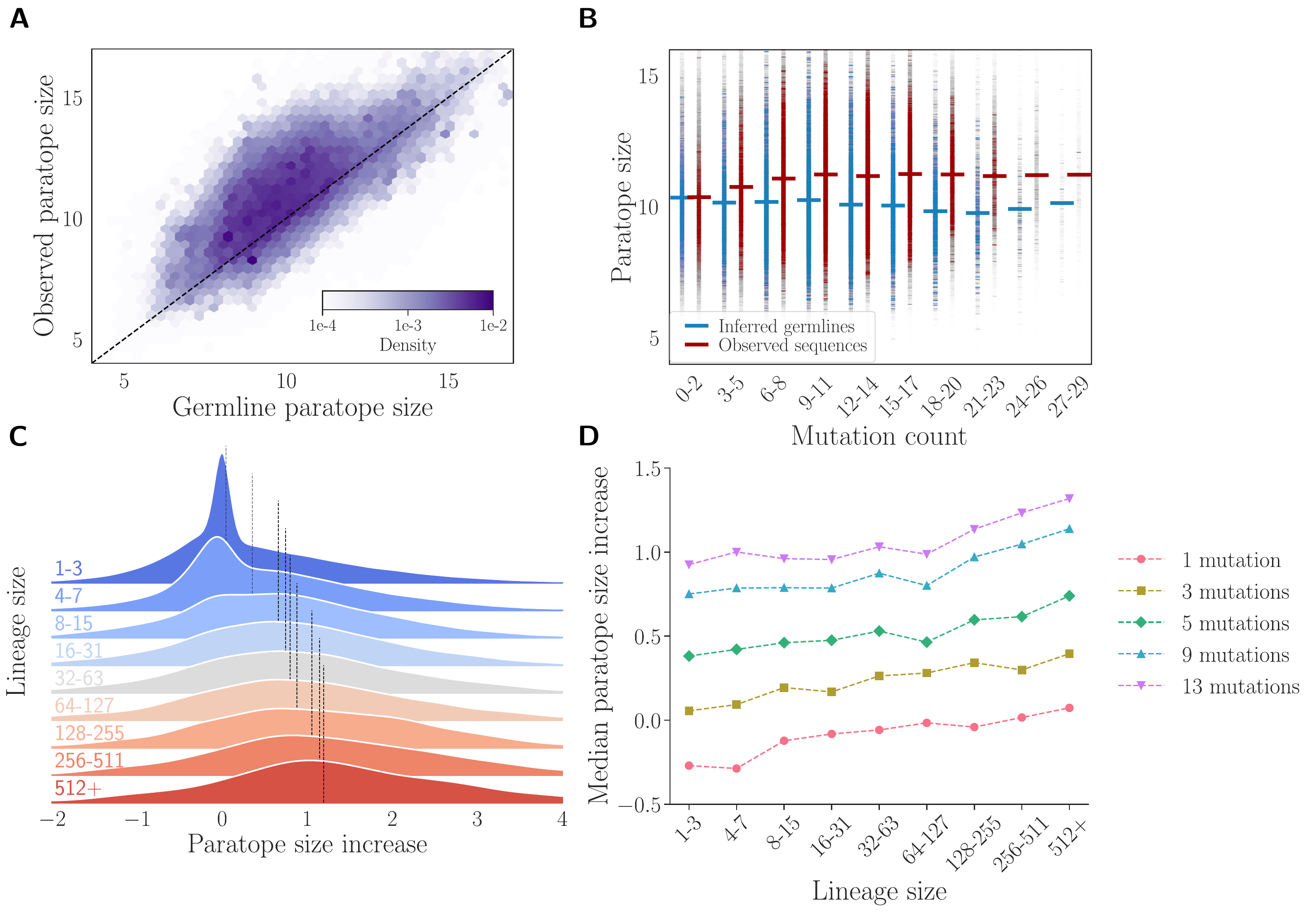}
	\caption{
		Effect of hypermutations on paratope size in human repertoires. Analysis for donor 326651 from \cite{briney2019commonality}.
		(A) 2D histogram showing the relationship between the paratope size of observed antibody sequences and their inferred germline counterparts.
		(B) Paratope sizes of observed sequences and germline sequences grouped by amino acid mutation count bins.
		(C) {\GA Density of the average increase in paratope size within lineages, shown across different lineage size bins. Each density curve is fitted using all lineages in the corresponding size range. The black line indicates the median average increase in paratope size for each bin.}
		(D) {\GA Median average increase when averaging over sequences with a fixed number of mutations within the lineage.}}
	\label{fig5}  
\end{figure*}
While the analysis of all intermediates between a naive and a matured antibody allows us to get a full picture of the sequence landscape for that particular pair, these sequences are not representative of actually explored variants in naturally occuring lineages found in antibody repertoires.
To address this limitation, we analyzed data from {Gerard et al.} \cite{gerard2020high} consisting of IgG {paired} heavy and light chain sequences from two mice immunized with tetanus toxoid (TT). Antigen-binding, IgG-expressing {B} cells were isolated using a {fluorescence-based droplet assay {\GA within} a microfluidic sorting system,} {\GA yielding 1,390 VH/VL pairs with $\sim$93\% of them binding to the tetanus toxoid (TT) antigen.
This resulted in a TT-immunized repertoire of TT-specific antibodies, which we compared to a naive antibody repertoire from mice of the same species reported by Goldstein et al. \cite{goldstein2019massively}. The naive repertoire was subsampled to match the heavy chain V gene family distribution observed in the TT-immunized repertoire.}
{\GA For each antibody repertoire, we inferred germline sequences, quantified hypermutation (SHM) counts, performed paratope predictions, and identified clonal lineages as detailed in Section~\ref{IV.F}.
As expected, the TT-binding repertoire exhibited significantly higher mutation counts compared to the naive repertoire (Fig.~\ref{si:fig5}). Additionally, the TT repertoire showed extensive clonal expansion, with 92\% of sequences belonging to multi-cell lineages, compared to only 10\% of sequences in naive mice, reflecting the different antigen exposure between the two repertoires.}
We found that antibodies from the {TT repertoire} exhibited larger paratope sizes compared to those from the naive repertoire (Fig.~\ref{fig4}C), suggesting that antigen-binding antibodies {contain} larger paratopes.
Additionally, we noted a significant increase in paratope size in the mutated sequences relative to their germline ancestors in both the naive and {TT} repertoires (Fig.~\ref{fig4}D). 
This {\GA increase} was particularly pronounced in the {TT} repertoire, where {we observed that} nearly all sequences had a larger paratope than their germline counterparts, suggesting that the process of {SHM that leads to affinity maturation occurs through the creation of larger paratopes that enhance antigen binding.}
Finally, we observed that in both repertoires, the paratope size  increased with the number of mutations in the hypermutated sequences (Fig.~\ref{fig4}E). 
{\GA Importantly, this effect was not observed in the germline sequences themselves, indicating that the increase in paratope size is a consequence of SHM rather than inherent differences in the original germline paratopes (Fig.\ref{fig4}E).}
{\GA Notably, the effect of somatic mutations on paratope size was more pronounced in the immunized (TT) repertoire, suggesting that the mutations observed in antigen-binding antibodies were preferentially selected to enlarge the paratope and enhance antigen recognition.}
However, the correlations between paratope size and mutation count in the two repertoires ($r = 0.12$, $p = 1 \times 10^{-5}$ for the TT repertoire; $r = 0.24$, $p = 1 \times 10^{-16}$ for the naive repertoire) are not directly comparable due to differences in their mutation count distributions.

To broaden our findings beyond the mouse immune system, {and to showcase the ability of Paraplume to be used for extremely large bulk repertoires}, we extended our analysis to a large healthy human antibody repertoire from Briney et al. \cite{briney2019commonality}. Because {\GA our model maintains strong performance on heavy chains even with single-chain inputs (Table~\ref{si:tab2})}, we applied the same methodology {\GA (cf. Section~\ref{IV.F})}
to this dataset, focusing specifically on IgG heavy chain sequences. After downloading the quality-processed reads {\GA from donor 316188, we retained approximately 4 million IgG sequences for analysis}.
Similarly to the mouse data, we  {found that the  observed (affinity-matured) human heavy-chain sequences had larger paratope sizes compared to their germline counterparts (Fig.~\ref{fig5}A) and the paratope size correlated positively  with the number of somatic mutations (Fig.~\ref{fig5}B, $r = 0.09$, $p < 10^{-16}$). However,  the paratope size plateaued for sequences with more than 10 mutations (Fig.~\ref{fig5}B)} {suggesting the possibility that additional mutations may be neutral{\TM, increase affinity within the paratope without affecting its size, or} work in a paratope-independent manner (e.g. by enhancing antibody stability)}. {\GA A similar early plateauing effect has also been reported in germinal center trajectory analyses, where most affinity gains occur within the first few mutations followed by a plateau and eventual decline, a phenomenon explained in part by survivorship biases \cite{dewitt2025replaying}}.

{\GA Building on our comparison between immunized and naive repertoires, which suggested that antigen-binding antibodies tend to have larger paratopes, we sought to explore the relationship between selection and paratope size within a human repertoire lacking antigen-specific sorting. To distinguish more strongly selected antibodies from less selected ones, we used clonal lineage size as a proxy for positive selection, assuming that larger lineages reflect more successful affinity maturation and proliferation.}
For each lineage, we computed the average increase in paratope size
of its sequences relative to their respective germlines. 
{\GA We observed that this average increase in paratope size positively correlates with lineage size (Fig.~\ref{fig5}C). 
Because larger lineages also tend to harbor more mutations, we controlled for mutation count and confirmed that the relationship between lineage size and paratope size increase remained robust (Fig.~\ref{fig5}D), indicating that paratope enlargement is associated with selection rather than mutation load alone.
Interestingly, sequences bearing only one or two mutations {\GA show a paratope size decrease} in small lineages (Fig.~\ref{fig5}D and Fig.~\ref{si:fig6}), suggesting that such mutations may have had deleterious effects on the paratope that limited clonal expansion.
Together, these findings suggest that paratope enlargement is more pronounced in lineages under stronger selection, likely reflecting additional rounds of affinity maturation and clonal expansion.}

\subsection{Paratope-Weighted Sequence Embeddings}
\begin{table*}[t]
	\scriptsize
	\centering
	\begin{tabular}{l|cc|cc|cc|cc}
		& \multicolumn{2}{c|}{\bf Binding affinity} & \multicolumn{2}{c|}{\bf Binding affinity} & \multicolumn{2}{c|}{\bf Binding affinity} & \multicolumn{2}{c}{\bf Expression} \\
		& \multicolumn{2}{c|}{\scriptsize (Shanehsazzadeh, $N=422$)} & \multicolumn{2}{c|}{\scriptsize (Warszawski, $N=2048$)} & \multicolumn{2}{c|}{\scriptsize (Koenig, $N=4275$)} & \multicolumn{2}{c}{\scriptsize (Koenig, $N=4275$)} \\
		\hline\hline
		{\bf PLM}& {\bf Paratope-weighted} & {\bf Unweighted} & {\bf Para-weighted} & {\bf Unweighted} & {\bf Para-weighted} & {\bf Unweighted} & {\bf Para-weighted} & {\bf Unweighted} \\
		\hline
		AbLang2 & \textbf{0.335} & 0.277 & 0.346 & \textbf{0.385} & 0.239 & \textbf{0.259} & 0.450 & \textbf{0.581} \\
		AntiBERTy & \textbf{0.313} & 0.289 & \textbf{0.259} & 0.169 & \textbf{0.236} & 0.197 & \textbf{0.439} & 0.358 \\
		ESM-2 & \textbf{0.312} & 0.307 & \textbf{0.380} & 0.334 & \textbf{0.311} & 0.305 & 0.656 & \textbf{0.678} \\
		IgT5 & 0.329 & \textbf{0.336} & 0.405 & \textbf{0.470} & 0.270 & \textbf{0.301} & 0.517 & \textbf{0.639} \\
		IgBert & \textbf{0.342} & 0.338 & 0.409 & \textbf{0.419} & \textbf{0.292} & 0.289 & 0.595 & \textbf{0.610} \\
		ProtT5 & \textbf{0.327} & 0.311 & \textbf{0.397} & 0.391 & \textbf{0.322} & 0.294 & 0.681 & \textbf{0.716} \\
		\hline
	\end{tabular}%
	\caption{Comparison of methods for generating sequence embeddings: The paratope-weighted embedding is computed as a weighted average of amino acid embeddings, with weights determined by their predicted  probabilities of belonging to a paratope, while the averaged embedding is a uniform mean across all amino acid embeddings. Performance is assessed using the $R^{2}$ score of a linear model predicting binding affinity or expression using the sequence embedding as input, across different protein language models (PLMs) and datasets.}
	\label{tab4}
\end{table*}
\begin{figure*}[t]
\includegraphics[width=\textwidth]{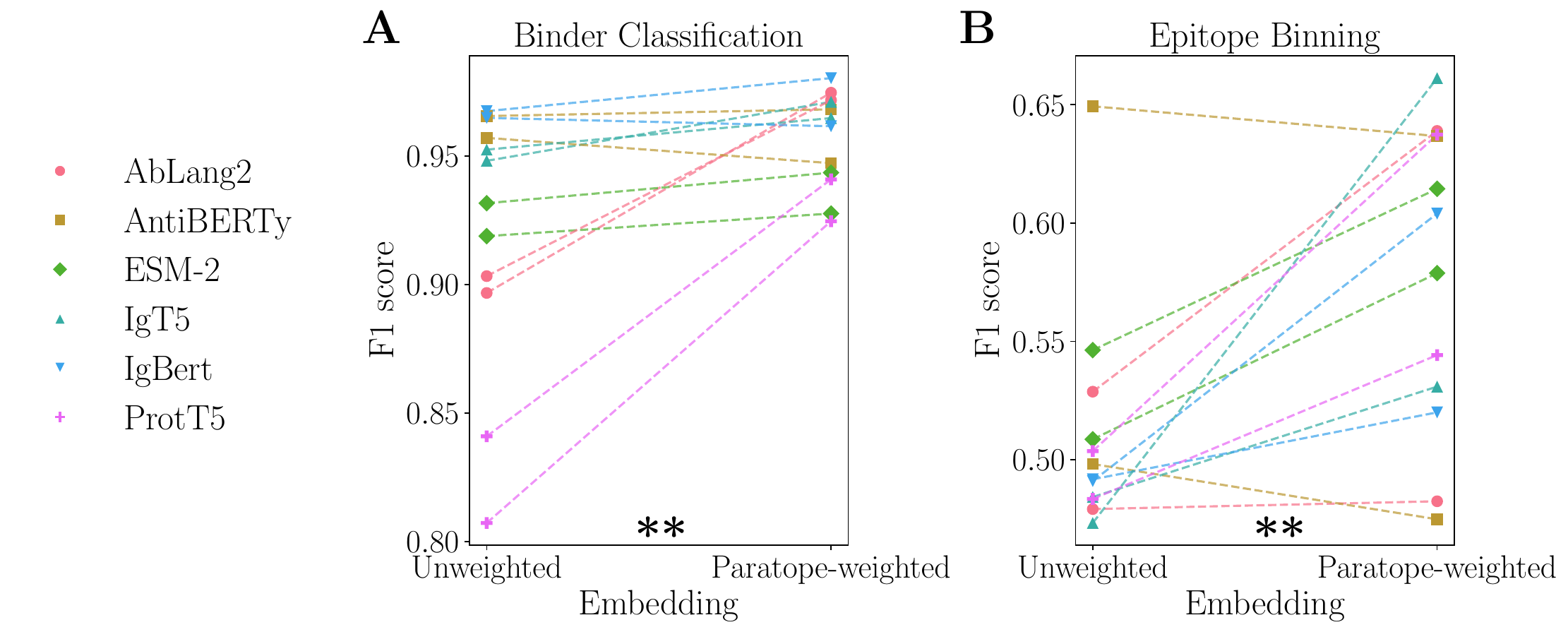}
\caption{(A) Comparison of the  paratope-weighted and {\GA unweighted} embeddings across the six large language models (LLMs) used in Paraplume. Performance is evaluated using the F1 score from a regression model trained to classify binders versus non-binders, based on sequences from \cite{phillips2021binding}. Two-fold cross-validation was performed on two distinct sets, resulting in the 12 data points.
(B) The same analysis as in (A), but with a regression model trained to classify antibodies into epitope classes using sequences from \cite{gowthaman2021cov3d}. A Wilcoxon paired sample test demonstrated that paratope-weighted embeddings yielded statistically significant improvements for both tasks, with p-values of 0.007 for binder classification and 0.004 for epitope binning.}
\label{fig6}  
\end{figure*}

We investigated whether incorporating paratope information could improve the prediction of binding affinity and enable classification of binders versus non-binders.  
Protein language model embeddings are widely used to generate fixed-dimensional sequence representations, which serve as input to neural networks that predict binding affinity. 
{\GA A common approach involves averaging the embeddings of all amino acids in a sequence, thereby treating all residues equally, regardless of whether they belong to the framework region, complementarity-determining regions (CDRs), or the paratope.}

Building on the work of {Ghanbarpour et al} \cite{ghanbarpour2023structure}, we propose a sequence representation that weights amino acid embeddings based on their probabilities of belonging to the paratope, calculated by Paraplume {\GA (cf. Section~\ref{IV.G})}. 
{\GA In our analysis, we compute and compare both unweighted and paratope-weighted representations for the six protein language models described Section~\ref{IV.A} and use them as inputs across multiple predictive tasks. For clarity, we refer to these as unweighted embeddings and paratope-weighted embeddings, respectively.} 

\subsubsection*{Affinity prediction}
{\GA We investigated whether incorporating paratope information improves binding affinity prediction for a linear regression model across three datasets with experimentally measured $K_D$ values (see Section~\ref{IV.H} for details of the experimental setup). 
Compared to unweighted embeddings, paratope-weighted embeddings yielded higher $R^2$ scores in the majority of cases (Table~\ref{tab4}), particularly when using protein language models not fine-tuned on antibody sequences such as ESM-2 and ProtT5. These results are consistent with the expectation that residues within the paratope contribute more strongly to binding affinity.

As a negative control, we repeated the analysis for the task of predicting antibody expression levels (cf. Section~\ref{IV.H}), a property for which paratope information should not be as informative. As expected, the paratope-weighted embeddings underperformed compared to unweighted embeddings (Table~\ref{tab4}), further supporting the specificity of paratope information for antigen-binding tasks.}
\subsubsection*{Antibody classification}
{\GA
We next tested the embeddings on two antibody classification tasks---distinguishing binders from non-binders, and predicting epitope specificity (epitope binning)---using datasets and logistic regression models described in Section~\ref{IV.I}. Across both tasks, paratope-weighted embeddings consistently outperformed unweighted embeddings (Fig.~\ref{fig6}), with again the most significant gains observed when using ESM, a PLM not fine-tuned on antibodies. These improvements were statistically significant, as confirmed by a Wilcoxon paired sample test ($p = 0.007$ for binder classification; $p = 0.004$ for epitope binning).}
  
\section{Discussion}

Mapping the specific location of the paratope is important for both biotechnology and medicine. In therapeutic antibody design, accurate identification of antigen-binding residues enables engineering of binding properties through targeted mutations. {Similarly, engineering therapeutic antibody developability often requires preserving paratope positions to avoid compromising binding function. Paratope residues are the most critical components of the antibody-antigen binding interface. Knowledge of how naive immunoglobulins evolve through the process of affinity maturation into effective antigen-specific antibodies, largely through expansion and change in paratope identity, is still poorly understood. Beyond individual antibodies, a rapid in-silico paratope prediction method holds great promise for the large-scale analysis of affinity maturation in antigen-specific antibody repertoires.} While sequencing technologies now allow high-throughput profiling of antibody repertoires, large-scale structural analysis remains challenging, as modeling thousands of antibodies is computationally demanding and often provides limited insight into the specific residues involved in antigen recognition.

Paratope prediction offers a scalable intermediate solution, bridging the gap between sequence-level data and functional interpretation. However, existing methods face several limitations that hinder their use in large-scale studies. Many rely on paired-chain inputs, restrict predictions to CDRs, or require structure prediction models, which limits throughput.  Paraplume's sequence-based and antigen-agnostic design offers a simpler and more scalable approach to studying mutational effects, eliminating the need for detailed structural modeling. We demonstrated that protein large language models can be used to develop a simple yet effective sequence-based paratope predictor. Paraplume avoids structural {input} dependencies, handles {both paired or} unpaired-chain data, and generalizes predictions across the full variable region. Despite its simplicity, Paraplume achieves performance on par with or exceeding that of current state-of-the-art sequence-based models on three different benchmark datasets.

To better understand the biological limits of sequence-based paratope prediction, we leveraged the symmetry of antibody arms to estimate the intrinsic variability in paratope usage. 
Our analysis shows that this variability reflects genuine biological differences rather than technical artifacts, and that the variability of an antibody's paratope is strongly correlated with that of its cognate antigen epitope. 
We were able to use this variability to define a realistic upper bound for prediction accuracy, offering a useful reference point for evaluating current and future predictive models.

{We further applied our model to investigate somatic hypermutation during antigen-driven immune responses and its influence on paratope identity. During affinity maturation within germinal centers, B cells undergo somatic hypermutation in the variable regions of both heavy and light chains of the B cell receptor. These mutations enhance the receptor’s affinity and specificity for the target antigen. B cells that successfully navigate iterative cycles of mutation, selection, and clonal expansion ultimately differentiate into plasma cells or memory B cells, expressing antibodies with improved binding characteristics.

  Our analysis reveals that affinity maturation in response to antigen exposure, which we measure through comparison of antigen-specific antibody sequences with their inferred germline state, is associated with an increase in predicted paratope size. This trend is particularly pronounced in clonally expanded antibody populations, indicating that enhanced antigen binding is a driving force behind this expansion. An expanded paratope allows for a greater number of chemically compatible interactions with the cognate epitope, thereby increasing binding affinity. Moreover, the requirement for additional interacting residues inherently demands a broader epitope interface, which in turn contributes to enhanced antibody specificity. This finding opens up the possibility of using changes in predicted paratope size as a proxy for increased antigen-specificity and affinity. This could be particularly useful for in silico methods of affinity maturation to predict those changes that will increase affinity and those that wont.
  
Another practical application of this work is the use of paratope-weighted embeddings whereby incorporating paratope information in PLM embeddings can enhance fine-tuning of models trained for antibody functional prediction. These paratope-weighted embeddings consistently outperform general averaged embeddings in prediction tasks associated with antibody function such as binder classification and affinity prediction.}
This work challenges the assumption that structural modeling is essential for studying antibody-antigen interactions and instead positions PLM-driven sequence-based paratope prediction as a powerful, scalable tool for repertoire-level analyses. 
In doing so, it opens new avenues for exploring the functional consequences of antibody diversification and evolution. 

Looking ahead, several components of our model stand to benefit from ongoing advancements. The continuous growth of structural antibody databases like SAbDab will enable training on larger and more diverse datasets. Simultaneously, improvements in protein language models, driven by increasing availability of sequence data and advances in representation learning, will enhance the quality of input embeddings.
Future work should aim to integrate these developments, with the goal of further improving paratope prediction accuracy and extending its applications in large-scale repertoire analysis{\AWO, therapeutic antibody affinity and developability engineering and generative antibody creation}.

\section{Methods} 

\subsection{Model Design Choices}
\label{IV.A}
{\GA
	\subsubsection*{Protein Large Language Models Embeddings}
	ESM-2 and ProtTrans are protein large language models (PLMs), whereas Antiberty is an antibody-specific model pretrained on 558M natural antibody sequences. IgT5 and IgBert are PLMs fine tuned on antibody sequences, and derive their names from the well-known NLP models T5~\cite{raffel2020exploring} and BERT~\cite{kenton2019bert}. One key difference between the two is that BERT predicts a single masked token at a time, whereas T5 does not have a predefined number of masked tokens to predict. To address the bias introduced by the predominance of germline-encoded residues in antibody sequences, Olsen et al.~\cite{olsen2024addressing} developed Ablang2, a model optimized for the prediction of mutated residues. 
	We assess the contribution of each of the six PLMs by comparing model performance under three settings: using all embeddings, using individual embeddings, and using all embeddings except one (Table~\ref{si:tab1}).
	Across the three benchmark datasets and four evaluation metrics, removing any single embedding led to a performance drop in at least one dataset, highlighting the complementarity of the six models.
	\subsubsection*{MLP architecture}
	The MLP architecture used in Paraplume comprises three hidden layers with dimensions 2000, 1000, and 500, respectively. 
	We incorporate several widely used regularization techniques such as dropout applied to the model weights, random masking of a portion of the input embeddings, and early stopping.
	We conducted a grid search on the Paragraph dataset to determine optimal hyperparameters, which were then used consistently across all three benchmark datasets. This approach avoids dataset-specific tuning and strengthens the robustness and generalizability of the model. A complete summary of the hyperparameter ranges explored and the final selected values is provided in Table~\ref{si:tab4}.
}

\subsection{Loss function}
\label{IV.B}
To train our model, we use the Binary Cross Entropy (BCE) loss function. It quantifies the difference between the model's predicted probability outputs and the true binary labels and is defined as:
\[ \text{BCE} = - \frac{1}{N} \sum_{i=1}^{N} \left( y_i \log(p_i) + (1 - y_i) \log(1 - p_i) \right), \]
where \( N \) is the number of samples, \( y_i \) is the true label for the \( i \)-th sample (either 0 or 1), and \( p_i \) is the predicted probability for the \( i \)-th sample. By minimizing the BCE loss during training, the model learns to output paratope probabilities that closely match the true labels for each amino acid, thereby improving its classification performance.
{\GA
\subsection{Single chain Vs Paired chain}
\label{IV.C}
Some PLMs are designed to process individual chains (ESM-2, ProtTrans, Antiberty), while others (IgBert, IgT5, AbLang2) are fine-tuned on paired antibody chains and can handle either paired or single chains. Paraplume supports both single-chain and paired-chain modes, depending on how the input embeddings are generated.
When both heavy and light chains are available, Paraplume generates embeddings by concatenating the two chains for models that operate on single sequences (e.g., ESM-2, ProtTrans, Antiberty), and passing each chain separately to models fine-tuned on paired chains (e.g., IgBert, IgT5, AbLang2).
In single-chain mode, only one chain (heavy or light) is used. Embeddings are computed using each PLM’s single-chain version, including those normally fine-tuned on paired data.
Thus, the distinction between the single and paired versions of Paraplume lies solely in how the embeddings are generated, not in the model architecture itself. In Table~\ref{si:tab2} we compare the two settings by evaluating Paraplume’s performance separately on heavy and light chains across the benchmark datasets.
}

\subsection{A dataset to analyze paratope asymmetry}
\label{IV.D}

To analyze paratope asymmetry we keep PDB files from the SabDab database~\cite{dunbar2014sabdab} meeting the following criteria, as shown Figure ~\ref{fig3}A: (1) the antigen must be a peptide or protein; (2) the antibody must consist of a heavy and light chain, thereby excluding nanobodies; (3) the antibody must have exactly 2 side chains; (4) for both the heavy and light chain, the Levenstein distance between the two side chains must be below 20, therefore reducing the risk to analyze antibodies engineered extensively to be bi-specific, or for which one of the side chains contains many missing residues. Following this {set of filters}, a total of 1,060 antibodies were retained for analysis. The metadata includes one row per antibody-antigen-interaction, describing one heavy and light chain of the antibody bound to an antigen chain.

\subsection{Paratope and epitope asymmetry}
\label{IV.E}

The paratope asymmetry between the paratopes of two identical antibody arms is defined as the count of all amino acids present in one of the two paratopes, but not in both.
Given two paratopes $P_1 = \{a_{\text{pos}_1}, \dots, a_{\text{pos}_n}\}$ and $P_2 = \{b_{\text{pos}_1}, \dots, b_{\text{pos}_m}\}$, where \( a_{\text{pos}_i} \) (respectively \( b_{\text{pos}_j} \)) represents the amino acid at position \( \text{pos}_i \) (\( \text{pos}_j \)) in the sequence, this can formally be written as a symmetric difference:
$$\text{Card}\left(\left(P_1 \cup P_2\right) \setminus \left(P_1 \cap P_2\right)\right)$$
For example, for two paratopes $\{L_{63}, Q_{64}, G_{66}\}$, $\{L_{63}, Q_{64}, A_{67}\}$ the asymmetry is $\text{Card}\left(\{G_{66},A_{67}\}\right)=2$
The normalized paratope asymmetry is then
\[
\begin{aligned}
	\frac{\text{Card}\left(\left(P_1 \cup P_2\right) \setminus \left(P_1 \cap P_2\right)\right)}{\text{Card}\left(P_1 \cup P_2\right)}
\end{aligned}.
\]
which corresponds to the Jaccard distance $d_J$ $$=1 - \frac{\text{Card}\left(P_1 \cap P_2\right)}{\text{Card}\left(P_1 \cup P_2\right)}= 1 - J(P_1, P_2)= d_J(P_1, P_2)$$
A high asymmetry is close to 1, whereas a low asymmetry is close to 0.
\newline
Epitope asymmetry and normalized epitope asymmetry are defined analogously using the epitopes of the two identical antigens bound to each of the antibody’s arms.

\subsection{Antibody repertoire analysis}
\label{IV.F}
{\GA Germline versions of each antibody were generated by identifying the closest V and J germline genes using IgBlast \cite{ye2013igblast}. The V and J regions of each antibody were then replaced with the inferred germline sequences, while retaining the original CDR3 sequences in both heavy and light chains due to the difficulty of accurately inferring germline CDR3 regions. This approach allowed for paratope prediction across the entire variable region. Somatic hypermutations (SHMs) were defined as the number of amino acid differences between the original antibody sequences and their corresponding inferred germline counterparts. Lineages were inferred using HILARy \cite{spisak2024combining}, which offers high precision and minimizes erroneous clustering of {antibodies coming from} distinct lineages.
We predicted the paratopes of all antibodies as well as their germline ancestors with Paraplume trained on the complete expanded dataset from \cite{chinery2023paragraph}}.
\newline
\subsection{Unweighted Vs Paratope-Weighted Embedding}
\label{IV.G}
Let a sequence of amino acids be represented as a set of embeddings:
\[
\mathbf{E} = \{\mathbf{e}_1, \mathbf{e}_2, \dots, \mathbf{e}_N\}, \quad \mathbf{e}_i \in \mathbb{R}^d,
\]
where \( \mathbf{e}_i \) is a \( d \)-dimensional embedding of the \( i \)-th amino acid in a sequence of length \( N \).

The standard approach for sequence representation is to compute the unweighted mean of all amino acid embeddings:
\[
\mathbf{e}_{\text{avg}} = \frac{1}{N} \sum_{i=1}^{N} \mathbf{e}_i.
\]

To integrate paratope information, we compute a weighted average of the amino acid embeddings, where the weights are derived from the normalized paratope probabilities \( p_i \), representing the likelihood that the \( i \)-th amino acid is part of the paratope, as predicted by Paraplume:
\[
\mathbf{e}_{\text{para}} = \sum_{i=1}^{N} w_i \mathbf{e}_i, \quad \text{where} \quad w_i = \frac{p_i}{\sum_{j=1}^{N} p_j}.
\]

\subsection{Binding Affinity Prediction}
\label{IV.H}
{\GA
We followed the methodology of Kenlay et al. \cite{kenlay2024large}, using three datasets from \cite{shanehsazzadeh2023unlocking}, \cite{warszawski2019optimizing}, and \cite{koenig2017mutational}, containing 422, 2048, and 4275 antibody sequences, respectively, each paired with $K_D$ measurements against a target antigen. For each dataset, we applied regularized linear regression to predict $\log(K_D)$ from either unweighted or paratope-weighted embeddings, using 10-fold cross-validation. Model performance was evaluated using the coefficient of determination $R^2$ on the test sets.
To validate task-specific relevance of paratope information, we applied the same method to predict antibody expression levels using data from \cite{koenig2017mutational}.}
\subsection{Antibody Classification}
\label{IV.I}
{\GA
For the binder classification task, we used data from Phillips et al. \cite{phillips2021binding}, selecting 111 high-affinity antibodies targeting the FluB strain and pairing each with a low-affinity mutant differing by one residue. We fit a logistic regression model using sequence embeddings (paratope-weighted or unweighted) to predict binder status, using cross-entropy loss and evaluating performance via F1-score.
For epitope binning, we curated a dataset from CoV3D \cite{gowthaman2021cov3d} comprising 329 antibodies targeting the SARS-CoV-2 RBD, grouped into four epitope classes. We applied a one-vs-rest logistic regression framework, training one binary classifier per epitope group. The average F1-score across classes was used to assess overall performance.
For both tasks, datasets were split into two equal, non-overlapping subsets. Two-fold cross-validation was performed within each subset, across six PLMs, resulting in 12 evaluations per task and embedding type. Results were averaged over five random seeds to ensure robustness. Statistical comparisons between embedding strategies were conducted using the Wilcoxon paired sample test.
}
\subsection{Data and code availability}
\label{IV.J}
{\GA Paraplume is freely available for non-commercial use as a PyPI package and can be accessed at https://github.com/statbiophys/Paraplume/. The package is designed for ease of use and includes the complete pipeline, covering dataset preparation and paratope labeling, model training, and model inference, thereby enabling full reproducibility of our results. Both Paraplume and its variant Paraplume-S support GPU and CPU execution, as well as single-chain and paired-chain inputs. Users can readily retrain Paraplume on larger datasets with customized parameter settings, including selection of subsets among the six PLMs employed in this work. Details and data of all benchmark and application experiments are provided in https://zenodo.org/records/17021232 to ensure reproducibility.}

\section*{Acknowledgements}
The study was supported by European Research Council Proof of Concept 101185627. A.W. is a bio-rad employee and may hold shares and/or stock options
in the company. We declare that this study received funding from
bio-rad. The funder collaborated directly in the study and was
involved in the study design, analysis, and interpretation of data,
the writing of this article, and the decision to submit it for
publication.

\nolinenumbers

\bibliographystyle{pnas}

\onecolumngrid
\renewcommand{\thefigure}{S\arabic{figure}}
\renewcommand{\thetable}{S\arabic{table}}
\setcounter{figure}{0}
\setcounter{table}{0}
\clearpage

\section{Supplementary information}

\begin{table}[H]
	\scriptsize
	\centering
	\begin{tabular}{l|cccc|cccc|cccc}
		\multicolumn{1}{c}{} & \multicolumn{4}{c}{\textbf{PECAN dataset}} & \multicolumn{4}{c}{\textbf{Paragraph dataset}} & \multicolumn{4}{c}{\textbf{MIPE dataset}} \\
		\cline{2-13}
		\textbf{PLM Embedding} & \textbf{PR} & \textbf{ROC} & \textbf{MCC} & \textbf{F1} & \textbf{PR} & \textbf{ROC} & \textbf{MCC} & \textbf{F1} & \textbf{PR} & \textbf{ROC} & \textbf{MCC} & \textbf{F1} \\
		\hline\hline
		\bf All (Paraplume)                    & 0.730 & \underline{0.963} & 0.657 & \underline{0.682} &\underline{0.758} & \bf 0.966 & \underline{0.676} & \underline{0.701} & 0.716 & \underline{0.962} & \bf 0.632 & \bf 0.651 \\
		All except AbLang2     & 0.729 & \underline{0.963} & 0.653 & 0.678 & 0.756 &\bf 0.966 & 0.674 & 0.699 & 0.716 & \bf{0.963} & 0.629 &  0.649 \\
		All except antiBERTy       & \bf 0.735 & \bf 0.964 & \bf 0.660 & \bf 0.685 & 0.756 & \bf 0.966 & 0.671 & 0.696 & 0.725 & \bf{0.963} & 0.626 & 0.647 \\
		All except IgBert    & 0.732 & \bf 0.964 & 0.656 & 0.681 & 0.757 & \bf 0.966 &\underline{0.676} & \underline{0.701} & 0.717 & \underline{0.962} & \underline{0.630} & \underline{0.650} \\
		All except IgT5   & 0.730 & \underline{0.963} & 0.657 & 0.681 &\bf  0.759 & \bf 0.966 & \bf 0.677 & \bf0.702 & 0.716 & \underline{0.962} & \underline{0.630} & 0.649 \\
		All except ProtT5      & 0.727 & \underline{0.963} & \underline{0.658} &\underline{0.682} & 0.755 & \bf 0.966 & 0.660 & 0.686 & 0.714 & \underline{0.962} & 0.629 & 0.648 \\
		All except ESM-2         & 0.730 & \underline{0.963} & 0.657 & \underline{0.682} & 0.754 &0.965 & 0.672 & 0.697 & 0.718 & \underline{0.962} & \underline{0.630} & 0.649 \\
		\hline
		\bf ESM-2 (Paraplume-S)               & \underline{0.734} & \underline{0.963} & 0.650 & 0.677 & 0.755 & \bf 0.966 & 0.660 & 0.686 &\underline{0.726} & \bf 0.963 & 0.629 & 0.649 \\
		AbLang2                                & 0.693 & 0.955 & 0.599 & 0.624 & 0.731 & 0.962 & 0.628 & 0.653 & 0.698 & 0.955 & 0.582 & 0.603 \\
		antiBERTy                              & 0.631 & 0.928 & 0.553 & 0.588 & \underline{0.682} & \underline{0.939} & \underline{0.574} & \underline{0.602} & 0.608 & 0.924 & 0.519 & 0.547 \\
		IgBert                                 & 0.710 & 0.958 & 0.596 & 0.620 & 0.739 & 0.963 & 0.632 & 0.657 & 0.696 & \underline{0.962} & 0.594 & 0.614 \\
		IgT5                                   & 0.703 & 0.959 & 0.574 & 0.596 & 0.737 & 0.963 & 0.642 & 0.667 & 0.688 & 0.959 & 0.574 & 0.594 \\
		ProtT5                                & \underline{0.734} & \underline{0.963} & 0.633 & 0.656 & 0.752 & 0.965 & 0.633 & 0.656 & \bf 0.730 & \bf{0.963} & 0.611 & 0.630 \\
		\hline
	\end{tabular}
\caption{Ablation study evaluating the impact of different PLM embedding configurations on performance across three datasets. The default Paraplume configuration uses all six embeddings: AbLang2, antiBERTy, IgBert, IgT5, ESM-2, and ProtT5. We report results when each embedding is removed individually or used alone. Bold indicates the best score, and underlined values represent the second-best. Based on performance across all 12 evaluation points, we chose to retain all six embeddings. While choosing different settings for each dataset could yield higher scores, we prioritize robustness and use the same configuration across all three datasets. Paraplume-S is a lightweight variant of Paraplume that uses only ESM-2 embeddings. All results are averaged over 16 random seeds to account for variability.}
	\label{si:tab1}
\end{table}

\begin{table}[H]
	\centering
	\scriptsize
	\begin{tabular}{cc|cccc|cccc|cccc}
		\multicolumn{2}{c|}{} & \multicolumn{4}{c}{\textbf{PECAN dataset}} & \multicolumn{4}{c}{\textbf{Paragraph dataset}} & \multicolumn{4}{c}{\textbf{MIPE dataset}} \\
		\cline{3-14}
		\textbf{Model} & \textbf{Test chain} & \textbf{AP} & \textbf{ROC} & \textbf{MCC} & \textbf{F1} & \textbf{AP} & \textbf{ROC} & \textbf{MCC} & \textbf{F1} & \textbf{AP} & \textbf{ROC} & \textbf{MCC} & \textbf{F1} \\
		\hline\hline
		Paraplume-Paired & Heavy & 0.771 & 0.966 & 0.666 & 0.690 & 0.794 & 0.969 & 0.701 & 0.724 & 0.726 & 0.959 & 0.623 & 0.645 \\
		Paraplume-Single & Heavy & 0.766 & 0.965 & 0.658 & 0.682 & 0.788 & 0.968 & 0.682 & 0.709 & 0.725 & 0.959 & 0.592 & 0.608 \\
		\hline
		Paraplume-Paired & Light & 0.749 & 0.967 & 0.607 & 0.617 & 0.753 & 0.969 & 0.639 & 0.647 & 0.758 & 0.972 & 0.638 & 0.644 \\
		Paraplume-Single & Light & 0.671 & 0.953 & 0.519 & 0.534 & 0.683 & 0.958 & 0.578 & 0.591 & 0.736 & 0.965 & 0.567 & 0.568 \\
	\end{tabular}
	\caption{Comparison of Paraplume using paired and single chain embeddings. We evaluate Paraplume's performance on individual chains (heavy or light, as indicated in the Test chain column) by comparing two input settings: embeddings generated from both chains (Paraplume-Paired) versus embeddings generated from only the test chain (Paraplume-Single). Results show that using heavy chain embeddings leads to a slight but acceptable drop in performance compared to using paired chain embeddings, indicating that Paraplume remains robust even in the absence of paired chain information.}
	\label{si:tab2}
      \end{table}

\begin{table}[H]
	\centering
	\scriptsize
	\begin{tabular}{l|ccc|ccc}
		\multirow{2}{*}{\bf Ab region} & \multicolumn{3}{c|}{\bf AP AUC} & \multicolumn{3}{c}{\bf ROC AUC} \\
		& {\bf Paraplume} & {\bf Paragraph-ABB} & {\bf Paragraph-crystal} & {\bf Paraplume} & {\bf Paragraph-ABB} & {\bf Paragraph-crystal} \\
		\hline\hline
		CDR1 light       & \underline{0.786} & 0.762 & \textbf{0.800} & 0.911 & 0.857 & \textbf{0.928} \\
		CDR2 light       & 0.451 & \textbf{0.675} & \underline{0.452} & \underline{0.990} & 0.876 & \bf{0.991} \\
		CDR3 light       & \underline{0.790} & 0.770 & \textbf{0.809} & \underline{0.941} & 0.884 & \textbf{0.953} \\
		\hline
		CDR1 heavy       & \underline{0.790} & 0.735 & \textbf{0.803} & \underline{0.923} & 0.856 & \textbf{0.934} \\
		CDR2 heavy       & \textbf{0.805} & 0.789 & \underline{0.804} & \textbf{0.931} & 0.854 & \underline{0.930} \\
		CDR3 heavy       & \underline{0.838} & 0.796 & \textbf{0.893} & \underline{0.893} & 0.866 & \textbf{0.922} \\
		\hline
		Framework        & \textbf{0.668} & 0.429 & \underline{0.566} & \textbf{0.977} & 0.768 & \underline{0.831} \\
	\end{tabular}
	\caption{Comparison of AP and ROC metrics for Paraplume and Paragraph across different regions of the antibody sequence. Paragraph-ABB refers to Paragraph using structures modeled with ABodyBuilder, while Paragraph-crystal refers to Paragraph trained on experimentally determined structures. Results for Paragraph-ABB are taken from the original study \cite{chinery2023paragraph}, whereas results for Paragraph-crystal were computed by retraining Paragraph on experimentally-determined structures.}
	\label{si:tab3}
      \end{table}

\begin{table}[H]
	\centering
	\scriptsize
	\begin{tabular}{c|c|c}
				\bf Hyperparameter       & \bf Range                                             & \bf Optimal Value(s) \\
				\hline\hline
				Embeddings               & \makecell[c]{igT5, antiberty, ablang2, \\ igbert, esm, prot-t5, all} & all \\
				\hline
				\makecell[c]{Dimensions of hidden \\ layers ($dim_1,...,dim_n$)} & \makecell[c]{(4000, 2000, 1000) \\ (2000, 1000, 500) \\ (4000, 2000, 1000, 500) \\ (2000, 1000, 500, 250)} & (2000, 1000, 500) \\
				\hline
				Learning Rates           & 1e-5, 5e-5                                           & 1e-5 \\
				Dropout Rates            & 0.2, 0.3, 0.4                                        & 0.4 \\
				Masking Probabilities    & 0, 0.4                                               & 0.4 \\
				Batch Sizes              & 8, 16, 32                                            & 16 \\
				
				L2 Penalties             & 0, 1e-5                                              & 1e-5 \\
				Imbalance weighting      & 1, 1.2                                               & 1 \\
			\end{tabular}
			\caption{Summary of hyperparameters explored, their ranges, and optimal values on the Paragraph dataset.}
			\label{si:tab4}
\end{table}

\begin{figure}[H]
	\includegraphics[width=0.95\textwidth]{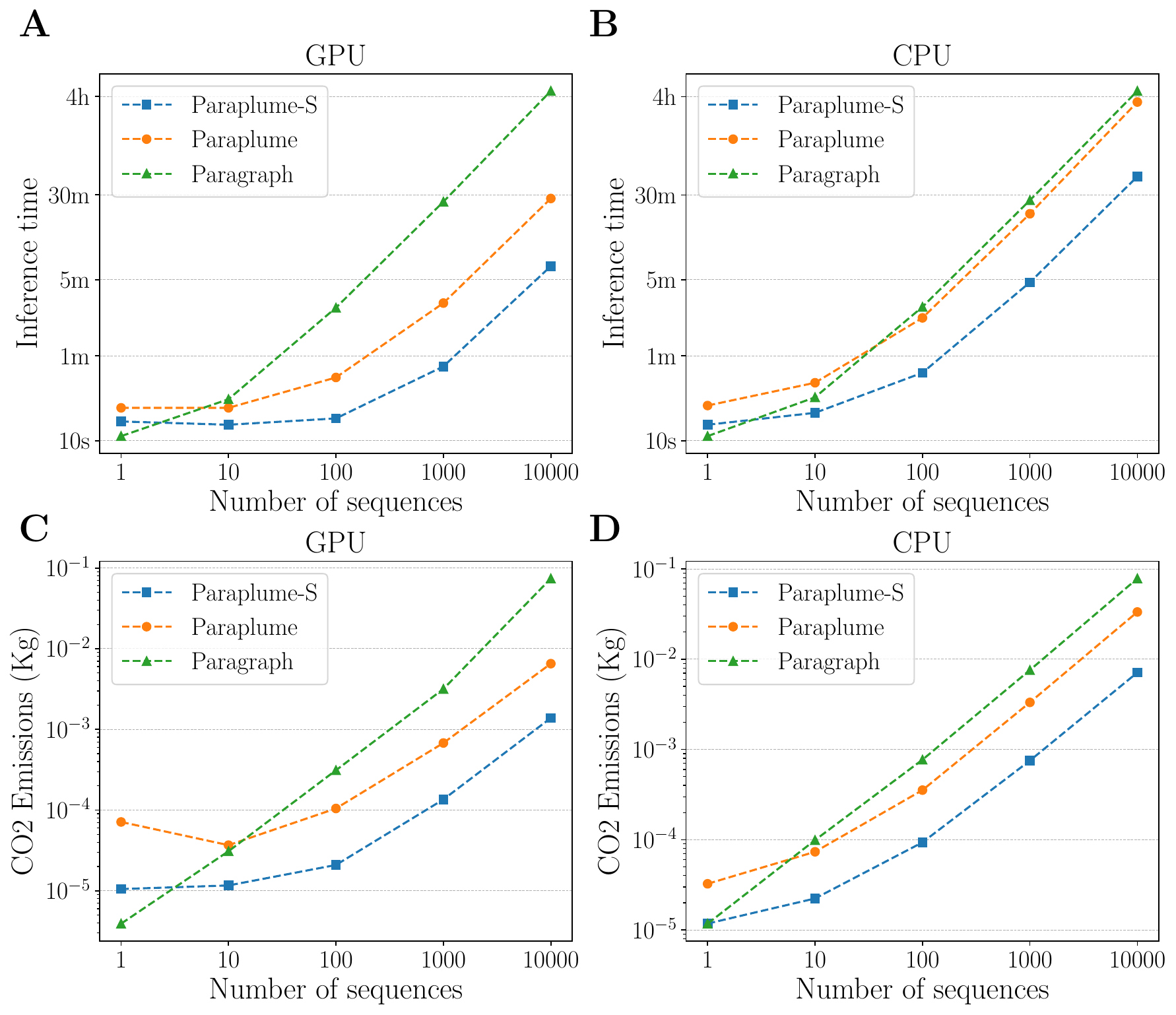}
	\caption{{\GA Comparison of inference time and $\text{CO}_2$ emissions for Paragraph, Paraplume, and Paraplume-S. Inference time was compared across different numbers of sequences on an NVIDIA RTX 5000 Ada Generation GPU (A) and 96-core Intel(R) Xeon(R) Gold 6442Y CPUs (B). For Paragraph, 3D structures were generated using AbodyBuilder3, the fastest available structure prediction tool to our knowledge, to ensure a fair comparison. We also compared $\text{CO}_2$ emissions using the package codecarbon \cite{benoit_courty_2024_11171501}, on GPU (C) and CPU (D).}}
	\label{si:fig1}  
\end{figure}
     
\begin{figure}[H]
	\hspace{-0.3cm}
	\includegraphics[width=1.03\textwidth]{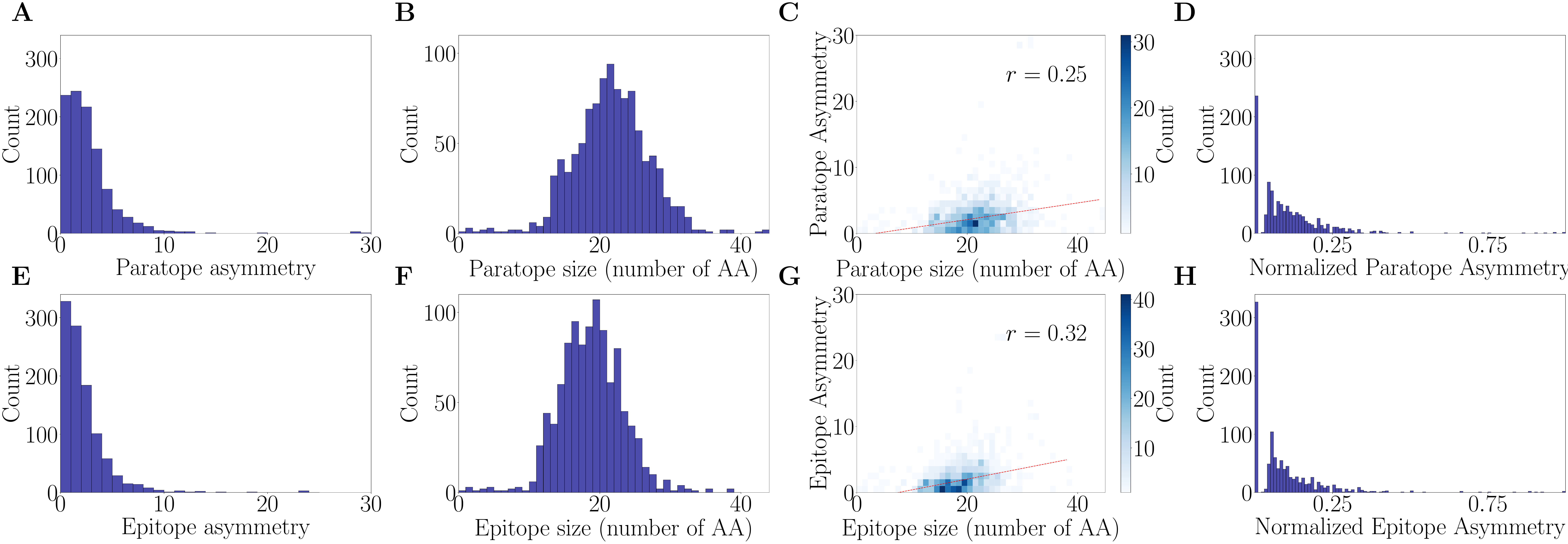}
	\caption{{\GA Statistics of the dataset curated to study paratope asymmetry (1039 antibody-antigen complexes)}. Histograms of the (A) paratope asymmetry and (B)
          paratope size. (C) Heatmap of paratope asymmetry against paratope size, colored by number of sequences. $r$ is the Pearson correlation coefficient.
           (D) Histogram of the paratope asymmetry normalized by the paratope size. (E-H) Same as (A-D) but for the epitope. 
	}
	\label{si:fig2}  
\end{figure}

\begin{figure}[H]
	\hspace{-0.3cm}
	\includegraphics[width=1.03\textwidth]{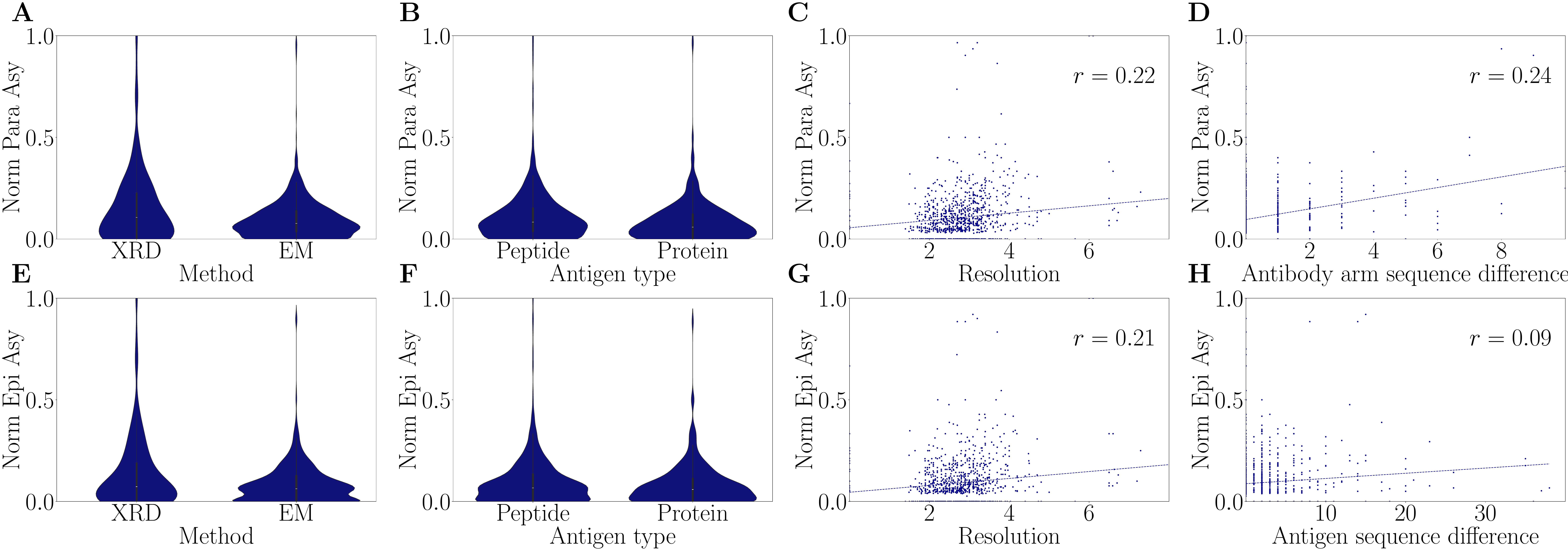}
	\caption{{\GA Paratope and epitope asymmetry against PDB characteristics.} Violin plots of the normalized paratope asymmetry separated by crystallography method (A) and antigen type (B). Normalized paratope asymmetry against PDB resolution (C) and Levenstein distance between the two antibody arms(D). (E-H) Same but for the normalized epitope asymmetry. 
	}
	\label{si:fig3}  
\end{figure}

\begin{figure}[H]
	\includegraphics[width=1\textwidth]{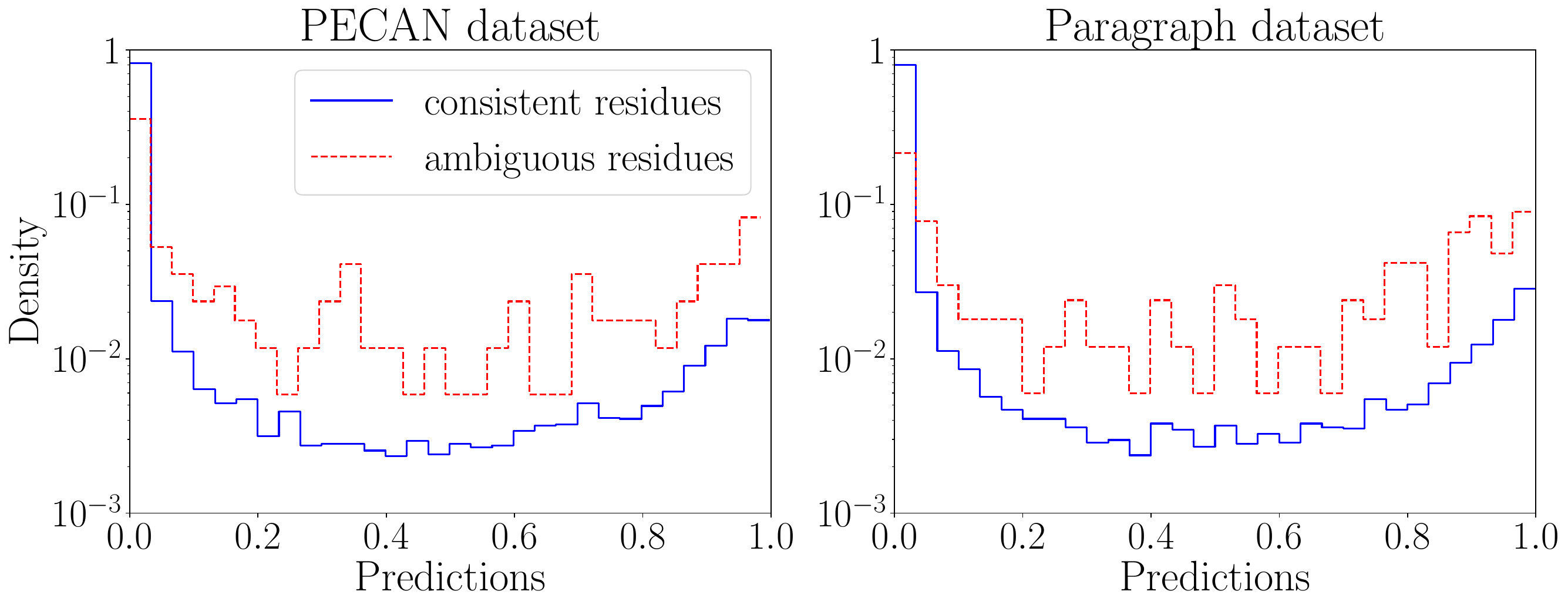}
	\caption{{\GA Comparison of Paraplume's predictions for consistent residues (same paratope label in both arms) and ambiguous residues (different paratope labels in both arms) in the PECAN dataset (left) and Paragraph dataset (right).}}
	\label{si:fig4} 
\end{figure}
     
\begin{figure}[H]
	\includegraphics[width=1\textwidth]{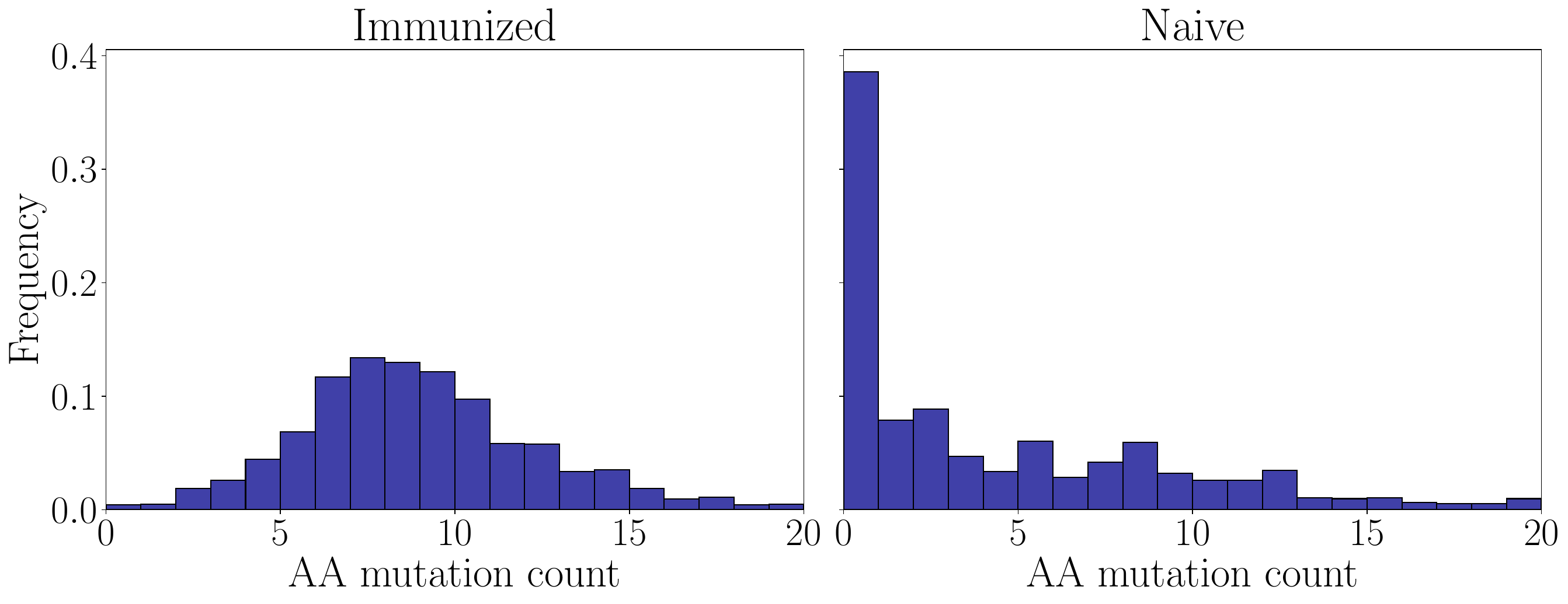}	\caption{{\GA Amino acid} mutation count distribution for (A) the immunized mouse antibody repertoire of \cite{gerard2020high} and (B) the naive mouse antibody repertoire of \cite{goldstein2019massively}.}
	\label{si:fig5} 
\end{figure}

\begin{figure}[H]
	\includegraphics[width=1\textwidth]{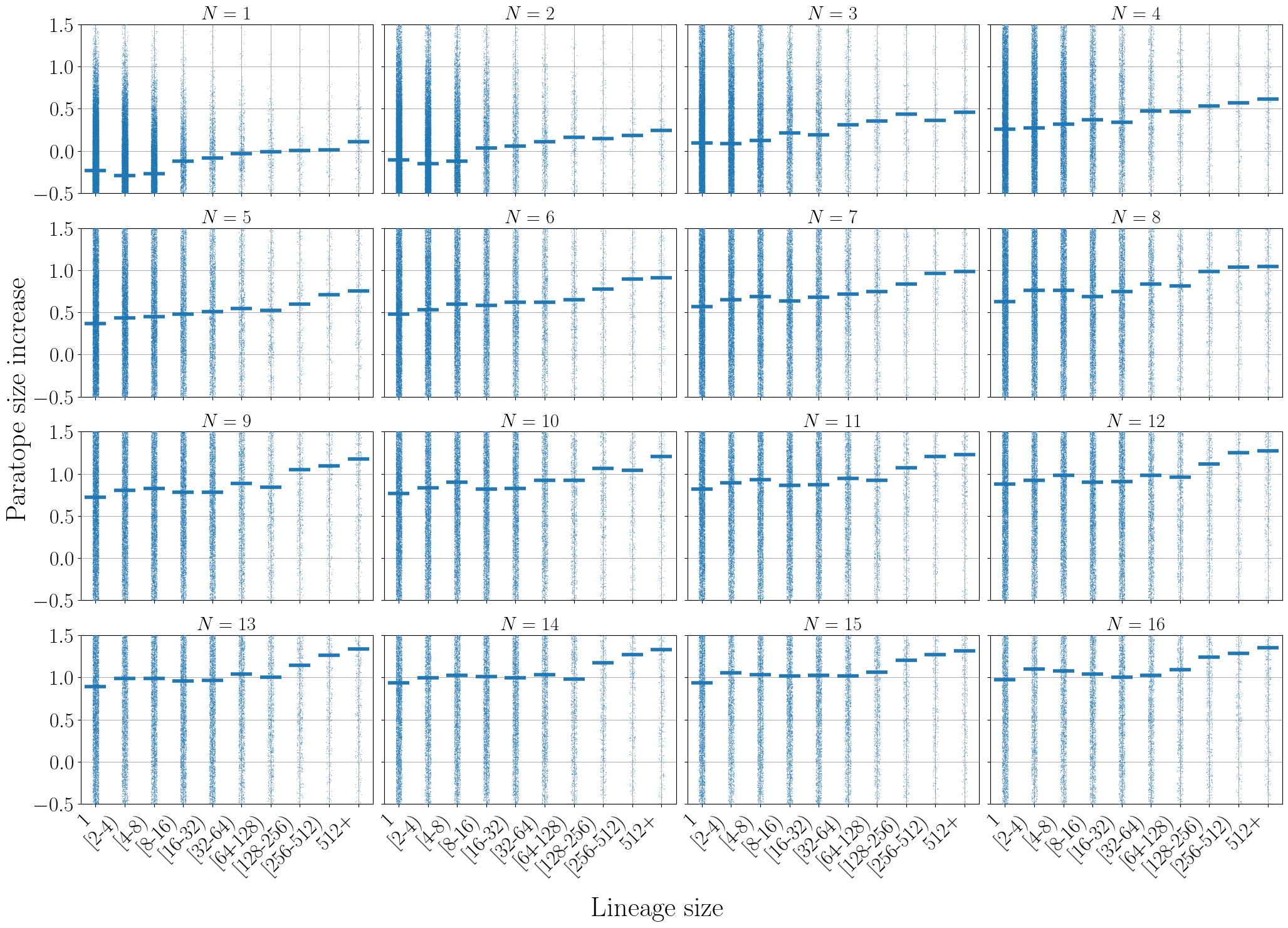}
	\caption{{\GA Average paratope size increase across different lineage sizes. The average in computed over sequences with a fixed number of mutations within the lineage (from $N=1$ top left to $N=16$ bottom right). Each point is a lineage, and the mean average increase is the thick line.}}
	\label{si:fig6}  
\end{figure}

\end{document}